\documentclass[11pt]{article}
\usepackage{amsmath,amssymb,color,graphics,epsfig,cite,ulem}

\usepackage{amsfonts}
\usepackage[unicode=false,
bookmarks=true,bookmarksnumbered=false,bookmarksopen=false,
breaklinks=false,pdfborder={0 0 1},backref=false,colorlinks=false]
{hyperref}

\newcommand{\hoch}[1]{$\, ^{#1}$}

\usepackage{comment}
\newcommand{\be}{\begin{equation}}
	\newcommand{\ee}{\end{equation}}
\newcommand{\bea}{\setlength\arraycolsep{2pt} \begin{eqnarray}}
	\newcommand{\eea}{\end{eqnarray}}
\newcommand{\nn}{\nonumber}

\def\ft#1#2{{\textstyle{\frac{\scriptstyle #1}{\scriptstyle #2} } }}
\def\fft#1#2{{\frac{#1}{#2}}}

\def\0{{\sst{(0)}}}
\def\1{{\sst{(1)}}}
\def\2{{\sst{(2)}}}
\def\3{{\sst{(3)}}}
\def\4{{\sst{(4)}}}
\def\5{{\sst{(5)}}}
\def\6{{\sst{(6)}}}
\def\7{{\sst{(7)}}}
\def\8{{\sst{(8)}}}
\def\sst#1{{\scriptscriptstyle #1}}

\def\a{{\alpha}}
\def\tr{\widetilde{r}}

\def\m{\mu}
\def\n{\nu}
\def\l{\lambda}
\def\r{\rho}

\begin{document}
	
	\begin{center}
		{\Large {\bf Improved Reall-Santos method for AdS black holes\\ in general 4-derivative gravities }}
		
		\vspace{20pt}
		
		{\large Peng-Ju Hu\hoch{1}, Liang Ma\hoch{1}, H. L\"u\hoch{1,2,3} and Yi Pang\hoch{1,3}}
		\vspace{10pt}
		
		{\it \hoch{1}Center for Joint Quantum Studies and Department of Physics,\\
			School of Science, Tianjin University, Tianjin 300350, China }
		
		{\it \hoch{2}Joint School of National University of Singapore and Tianjin University,\\
			International Campus of Tianjin University, Binhai New City, Fuzhou 350207, China}
		
		{\it \hoch{3}Peng Huanwu Center for Fundamental Theory, Hefei, Anhui 230026, China}
		
		\vspace{40pt}
		
		\underline{ABSTRACT}
	\end{center}
	
	For asymptotically flat  black holes, Reall-Santos method is a convenient tool to compute leading higher derivative corrections to the thermodynamic quantities  without actually solving the modified field equations. However, there are subtleties in its generalization to asymptotically AdS black holes with general higher derivative corrections. First of all, it is necessary to know all the higher derivative holographic counterterms and the surface terms implementing the variational principle and subtracting the divergence. One then needs to solve for the modified AdS radius and rescale the time coordinate in an appropriate way such that the induced metric on the conformal boundary of AdS black hole is not modified. We observe that  Reall-Santos method can be directly applied to a particular 4-derivative gravity model, known as the Einstein-Weyl gravity, which does not modify the AdS radius and requires only the Gibbons-Hawking-York term and holographic counterterms for the 2-derivative theory. We thus suggest that to compute the thermodynamic quantities of AdS black holes in general 4-derivative theories of gravity, one simply needs to transform it to a Einstein-Weyl gravity with identical thermodynamic variables by appropriate field redefinitions. We explicitly verify this proposal with spherically-symmetric and static charged black holes in Einstein-Maxwell theory extended with generic 4-derivative interactions.
	


	\vfill {\footnotesize pengjuhu@tju.edu.cn\ \ \  liangma@tju.edu.cn \ \ \ mrhonglu@gmail.com\ \ \ pangyi1@tju.edu.cn}
	
	
	\thispagestyle{empty}
	\pagebreak

	\tableofcontents
	\addtocontents{toc}{\protect\setcounter{tocdepth}{2}}
	
	\newpage
	\section{Introduction}
	\label{sec:intro}
	
	Black holes, branes and stars that are asymptotic to anti-de Sitter (AdS) spacetimes have played a very important role in the AdS/CFT correspondence. In this framework, the phase structure of the dual CFTs is encoded in the thermodynamics of various solutions in bulk AdS that exhibit certain universal behaviors at the leading order of Maldacena limit. For instance, the entropy obeys an area law. The large charge limit of the mass of extremal black holes or boson stars exhibits the same behavior as the scaling dimension of the lowest operator with a large global U(1) charge \cite{Hellerman:2015nra, Liu:2020uaz,Guo:2020bqz}. Beyond the leading order, the precise comparison between the CFT and gravity results is still under development (see \cite{ Melo:2020amq,  Bobev:2022bjm, Cassani:2022lrk} for some recent attempts,) mainly due to the technical difficulties. While in the CFT side, integrability and localization techniques allow one to compute certain supersymmetric observables  beyond the large $N$ limit which ought to be compared to higher order corrections in the gravity side, analogs of these powerful tools are still lacking in the gravity side. In most cases, solving complicated gravitational field equations order by order in derivative expansions is still the only feasible method.
	
	For asymptotically flat black holes,  Reall and Santos (RS) \cite{Reall:2019sah} observed that the leading higher derivative corrections to thermodynamic variables can be obtained without actually solving the modified field equations.  One simply needs to evaluate the total action on the solution of the 2-derivative theory.\footnote{It is worth pointing out the RS method does not address the issue that in higher-dimensional extremal rotating black holes, the near horizon geometry under higher-order curvature perturbations can involve irrational power expansions, which makes the horizon geodesically impenetrable, giving rise to a new type of singularity on the horizon without curvature divergence \cite{Mao:2023qxq}.} The RS method has been generalized to higher order in \cite{Ma:2023qqj} which showed that perturbative solutions up to $(n-1)$-th order are sufficient to produce $n$-th order corrections to thermodynamic variables.   Recent works \cite{Melo:2020amq,Bobev:2022bjm,Cassani:2022lrk, Cassani:2023vsa} applied the RS method, assuming its  validity, to asymptotically AdS black holes in gauged supergravities with higher derivative corrections. However, as was pointed in \cite{Ma:2023qqj}, there are subtleties in the direct application of the RS method to asymptotically-AdS black holes. In this work, using the widely studied example of Einstein-Gauss-Bonnet (EGB) gravity, we shall first show that the naive application of RS method to AdS black holes yields an incorrect on-shell Euclidean action.
		A careful study shows that the result obtained by evaluating the action of the uncorrected solution  differs from the on-shell action of the solution with leading order higher derivative correction by the term \cite{Reall:2019sah, Cassani:2022lrk}
		\be
		\Delta I_E\propto \int_{\partial\mathcal{M}|_{r={\infty}}}d^{D-1}x\sqrt{h}T^{(0)ab}\delta_1 g_{ab}\, ,
		\label{Idiff}
		\ee
		where $T^{(0)ab}$ is the Brown-York stress tensor for the AdS black hole solution in the 2-derivative theory and $\delta_1 g_{ab}$ denotes the perturbations generated by the leading higher derivative terms. The resolution of the discrepancy relies on a proper rescaling of the time coordinate depending on the ratio between effective and bare AdS radius. Consequently, in terms of the new coordinates, the induced metric on the conformal boundary of the AdS black hole is not modified by higher derivative terms which amounts to set $\delta_1 g_{ab}=0$. Thus the original proposal of RS for AdS black holes (see \cite[footnote 6]{Reall:2019sah}) could give the correct answer as long as the induced metric on the conformal boundary is preserved. However, practically one needs to first solve for the modified AdS radius and obtain the explicit form of the surface terms associated with the bulk higher derivative interactions. Furthermore, it is necessary to work out all the holographic counterterms, which can be a great challenge for general higher-derivative bulk terms.  We can thus ask the question: can we improve the RS method for AdS black holes without having to construct both the surface terms and counterterms associated with higher derivative bulk interactions, as in the case of asymptotic flat black holes? As we will see later, this can indeed be achieved.

	It is recalled that in the case of asymptotically flat black holes, employing the RS method, one can prove that the thermodynamic variables are manifestly invariant under field redefinitions \cite{Ma:2023qqj} as they do not depend on parameters affected by field redefinitions. In the AdS case, a direct computation shows that various physical quantities can depend on parameters that transform under field redefinitions. Indeed, this has been observed in \cite{Liu:2008kt} which also failed to express physical quantities in terms of field-redefinition invariant combinations of the parameters. However, it is generally believed that physical quantities should be invariant under field redefinitions. To understand this issue, we consider examples including pure gravity and Einstein-Maxwell theory extended with generic 4-derivative interactions. It turns out that the key to the invariance of thermodynamic variables requires a proper rescaling of the metric, so that the standard asymptotic structure of the AdS black hole at infinity and the time-like Killing vector both remain the same. In the case of charged AdS black holes, if the field redefinition is performed directly on the thermodynamic variables, their invariance becomes evident after utilizing the freedom of shifting the integration constants in the solution by terms depending on  parameters in the field redefinition. This also explains why in general the physical quantities cannot be expressed solely in terms of field-redefinition invariant combinations of the parameters in the Lagrangian except for the case of pure AdS vacuum \cite{Li:2021jfh}.
	
	Having understood under which circumstance field redefinitions do not modify the thermodynamic variables, one is allowed to choose any equivalent field frame in which the computation can be performed. We find that the most convenient frame is given by the Einstein-Weyl gravity with a cosmological constant.  A general 4-derivative extension of Einstein gravity can always be brought into this form where the effective Newton's constant and cosmological constant encode information about 4-derivative couplings of the original theory. We find that for Einstein-Weyl gravity, corrections to the solution of the 2-derivative theory always fall off fast enough near the AdS boundary, since the Weyl term does not affect the effective cosmological constant. Consequently the RS method yields the same answer as the ordinary method requiring the knowledge of the perturbative solution.  However, we emphasize here that for Einstein-Weyl gravity, the surface terms on the AdS boundary are those of Einstein gravity only when the integration of the Weyl-squared term is finite by itself. Finally, we recast the results in terms of the coefficients in the original theory. By this way,  we obtain thermodynamics for Einstein-Maxwell theory with general 4-derivative couplings without actually solving the corrected field equations. Surprisingly, this method also produces the right Casimir energy without invoking any surface terms associated with four derivative interactions.
	
	The paper is organized as follows. In section \ref{sec:naive}, we use the example of Einstein-Gauss-Bonnet gravity to show that for the original RS method to be vaild, one  needs to solve for the modified AdS radius and rescale the time coordinate appropriately. In section \ref{sec:red}, we derive the conditions under which theories before and after field redefinitions possess equivalent thermodynamic variables. In section \ref{sec:improved}, we then propose an improvement of RS method for models with general 4-derivative interactions without the need of solving for the modified AdS radius and knowledge of higher derivative surface terms.  Firstly, we show that improved RS method applies directly to Einstein-Weyl gravity. Then utilizing appropriate field redefinitions, we can always relate a model with generic 4-derivative interactions to the Einstein-Weyl gravity with identical thermodynamic variables. We conclude the paper in section \ref{sec:con}. In appendices, we give the leading-order perturbative solutions and explicit corrections to the thermodynamic variables. We also present the mass-charge and entropy-charge relations in the extremal limit.

	\section{The original RS method for AdS black holes}
	\label{sec:naive}
	
	In this section, we will first obtain thermodynamic variables of asymptotically AdS black holes in EGB gravity in 5, 6 and 7 dimensions by directly evaluating the on-shell action of the corrected solution up to first order in the coefficient of the
	Gauss-Bonnet term. We then show in detail how this result can be reproduced by carefully applying the procedure suggested by Reall-Santos \cite{Reall:2019sah}. It turns out that before performing perturbative expansion, one needs to rescale the time coordinate so that the metric on the conformal boundary is not modified by the GB term at fixed temperature. Through this paper, we use $D$ to denote the spacetime dimensions ranging from 5 to 7.
	
	To proceed, we start from the action below
	\be
	I_{\rm{EGB}}=  I_{\mathrm{bulk}}+I_{\mathrm{GHY}}-I_{\text{ct}},
	\ee
	in which
	\bea
	I_{\rm{bulk}}&=&\frac{1}{16\pi G_{D}}\int_{\mathcal{M}}d^{D}x\sqrt{-g}\Big[R+
	\frac{(D-1)(D-2)}{\ell_{0}^{2}}+\alpha(R^{2}-4R^{\mu\nu}R_{\mu\nu}+
	R^{\mu\nu\rho\sigma}R_{\mu\nu\rho\sigma})\Big],
	\nn\\
	I_{\rm GHY}&=&\frac{1}{8\pi G_{D}}\int_{\partial\mathcal{M}}d^{D-1}x\sqrt{-h}\Big[K-
	2\alpha\Big(2\mathcal{G}_{ab}K^{ab}+\frac{1}{3}
	(K^{3}-3KK^{ab}K_{ab}+2K_{b}^{a}K_{c}^{b}K_{a}^{c})\Big)\Big],\label{EGBlag}
	\eea
	where $a,b,c,...$ label indices of coordinate on the boundary, $K_{ab}$ denotes the extrinsic curvature and $\mathcal{G}_{ab}$ is the Einstein tensor associate with the induced metric $h_{ab}$ on the boundary. The counterterms on the AdS boundary take the form \cite{Emparan:1999pm, Liu:2008zf,Brihaye:2008xu}
	\bea
	I_{\text{ct}} & =&\frac{1}{16\pi G_{D}}\int d^{D-1}x\sqrt{-h}\mathcal{L}_{\text{ct}}\,\quad\mathcal{L}_{\text{ct}}=
	\mathcal{L}_{(0)}+\mathcal{L}_{(2)}+\mathcal{L}_{(4)}+\cdots\,,
	\nn\\
	{\cal L}_{(0)}&=&\frac{2(D-2)}{\ell_{0}}\left(1-
	\frac{\alpha}{6\ell_{0}^{2}}(D-3)(D-4)\right),
	\nn\\
	{\cal L}_{(2)}&=&\frac{\ell_{0}}{(D-3)}\left(1+
	\frac{3\alpha}{2\ell_{0}^{2}}(D-3)(D-4)\right)\mathcal{R}\,,
	\nn \\
	{\cal L}_{(4)}&=&\frac{\ell_{0}^{3}\left(1-
		\frac{7\alpha}{2\ell_{0}^{2}}(D-3)(D-4)\right)}{(D-3)^{2}(D-5)}
	(\mathcal{R}^{ab}\mathcal{R}_{ab}-\frac{D-1}{4(D-2)}R^{2})
	+\frac{\alpha\ell_{0}}{(D-5)}\mathcal{L}_{\rm GB}(h)\,,
	\label{Lct}
	\eea
	where the counterterm ${\cal L}_{(4)}$ is needed in $D>5$. The spherically symmetric black hole solutions in EGB model is well known \cite{Cai:2001dz,Cvetic:2001bk}. Treating the Gauss-Bonnet coupling $\a$ as a small parameter, one obtains the solution up to and including first order in $\alpha$ as
	\bea
	ds_{D}^{2} & =&-h(r)dt^{2}+\frac{dr^{2}}{f(r)}+r^{2}d\Omega_{D-2,k}^{2}\,,\qquad h=h_{0}+\Delta h\,,\qquad f=f_{0}+\Delta f\,,\nonumber \\
	h_{0} & =&f_{0}=\frac{r^{2}}{\ell_{0}^{2}}+k-\frac{2\mu}{r^{D-3}}\,,
	\qquad\Delta h=\Delta f=\alpha(D-3)(D-4)\big(\frac{4\mu^{2}}{r^{2(D-2)}}+
	\frac{r^{2}}{\ell_{0}^{4}}\big)+\mathcal{O}(\alpha^{2})\,,
	\label{EGBsol}
	\eea
	where $k$ can be 1, 0 or $-1$. The black hole horizon is located at
	\begin{equation}
		r_{h}=r_{0}+\Delta r\,,\qquad \Delta r=-\alpha\frac{(D-4)(D-3)\left(k^{2}\ell_{0}^{4}+2kr_{0}^{2}\ell_{0}^{2}
			+2r_{0}^{4}\right)}{(D-3)kr_{0}\ell_{0}^{4}+(D-1)r_{0}^{3}\ell_{0}^{2}}\,,
	\end{equation}
	where $r_0$ denotes the horizon radius of the uncorrected solution and  $\mu$ has been replaced via $\mu=\frac{r_{0}^{D-3} \left(k\ell_{0}^{2}+ r_{0}^{2}\right)}{ 2\ell_{0}^{2}}$. The mass, temperature and entropy of the solution can be obtained using standard method. Up to first order in $\alpha$, the temperature is given by
	\bea
	T&=&\frac{D-3}{4 \pi  r_{0}}\big(k+\frac{(D-1) r_{0}^2}{(D-3)
		\ell_{0}^2}\big)+\Delta T\,,
	\nn\\
	\Delta T&= & -\alpha\frac{(D-4)(D-3)\left((D-3)(D-2)k^{3}\ell_{0}^{2}
		+(D(3D-11)+12)k^{2}r_{0}^{2}\right)}{4\pi r_{0}^{3}\left((D-3)k \ell_{0}^{2} +(D-1)r_{0}^{2}\right)}
	\nn\\
	&&-\alpha\frac{(D-4)(D-3)\left((2(D-2)D+6)kr_{0}\ell_{0}^{2} +2(D-1)r_{0}^{3}\right)}{4\pi\ell_{0}^{4}\left((D-3)k\ell_{0}^{2} +(D-1)r_{0}^{2}\right)}\,.
	\eea
	For later convenience, we also redefine the parameter $r_0$ as $r_0=  \tr_0+\delta \tr_0$, with
	\bea
	\delta \tr_{0}&= & -\alpha\frac{(D-4)(D-3)\left((D-3)(D-2)k^{3}
		\ell_{0}^{4}+(D(3D-11)+12)k^{2} \tr_{0}^{2}\ell_{0}^{2}\right)}{(D-3)^{2}k^{2} \tr_{0}\ell_{0}^{4}-(D-1)^{2}\tr_{0}^{5}} \nn \\
	&& -\alpha\frac{(D-4)(D-3)\left(2((D-2)D+3)k\tr_{0}^{3} \ell_{0}^{2}+2(D-1)\tr_{0}^{5}\right)}{(D-3)^{2}k^{2} \ell_{0}^{6}-(D-1)^{2}\tr_{0}^{4}\ell_{0}^{2}}\,,\label{dr0}
	\eea
	so that in terms of the new parameter $\tr_0$, the temperature does not depend on $\a$ explicitly. In this parametrization, we list all the thermodynamic variables below
	\bea
	T&=&\frac{D-3}{4 \pi  \tr_{0}}\Big(k+\frac{(D-1) \tr_{0}^2}{(D-3) \ell_{0}^2}\Big)\ ,
	\\
	M &=&\frac{(D-2)\tr_{0}^{D-3}\left(k\ell_{0}^{2}+\tr_{0}^{2}\right) \omega_{D-2,k}}{16\pi\ell_{0}^{2}G_{D}}+\Delta{M}\,,\label{MAdS}
	\nn\\
	\Delta{M}&= & -\frac{\alpha(D-4)(D-3)(D-2)k\tr_{0}^{D-5}\left(3(D-2)(D-1)k\tr_{0}^{2} \ell_{0}^{2}+2(D-1)^{2}\tr_{0}^{4}\right) \omega_{D-2,k}}{16\pi\ell_{0}^{2}G_{D} \left((D-3)k\ell_{0}^{2}-(D-1)\tr_{0}^{2}\right)}\nn \\
	&& -\frac{\alpha(D-4)(D-3)^{2}(D-2)^{2}k^{3}\ell_{0}^{2} \tr_{0}^{D-5}\omega_{D-2,k}}{16\pi G_{D}\left((D-3)k\ell_{0}^{2}-(D-1)\tr_{0}^{2}\right)}\,,
	\\
	{S}&=&\frac{\omega_{D-2,k}\tr_{0}^{D-2}}{4G_{D}}+\Delta{S}\,,\label{TSAdS}
	\nn\\
	\Delta{S} &=&-\alpha\frac{(D-3)(D-2)k\tr_{0}^{D-4}\left((D-5)(D-2)k \ell_{0}^{2}+2(D-3)(D-1)\tr_{0}^{2}\right) \omega_{D-2,k}}{4G_{D}\left((D-3)k\ell_{0}^{2}-(D-1)\tr_{0}^{2}\right)}\,,
	\label{RedefQuan}
	\eea
	where $\omega_{D-2,k}$ is the area $\omega_{D-2,k}=\int d\Omega_{D-2,k}$. In the case of $k=0,\,-1$, it can be regularized by introducing a cutoff.  From the mass, temperature and entropy we obtain the Helmholtz free energy
	\bea
	F&=&M-TS=\frac{\tr_{0}^{D-3}\left(k\ell_{0}^{2}-\tr_{0}^{2}\right) \omega_{D-2,k}}{16\pi\ell_{0}^{2}G_{D}}
	\nn\\
	&&\quad-\alpha\frac{(D-3)(D-2)k\tr_{0}^{D-5} \left((D-2)k\ell_{0}^{2}+2(D-1)\tr_{0}^{2}\right) \omega_{D-2,k}}{16\pi\ell_{0}^{2}G_{D}}\,.
	\eea
	Substituting the solution \eqref{EGBsol}  with redefined parameter $\tr_0$ into the action \eqref{Lct}, we find that the Euclidean action $I_E$ is given by
	\be
	TI_E= F+{\rm mod}[D,2]\times a_1\,,
	\label{IBR}
	\ee
	where the holographic Casimir energy  $a_1$ appears in odd spacetime dimensions and takes the form \cite{Emparan:1999pm,Brihaye:2008xu}
	\bea
	a_1&=&\frac{[(D-2)\text{!!}]^{2}(-k)^{\frac{D-1}{2}}}{(D-1)!\ell_{0}} \frac{\omega_{D-2,k}}{8\pi G_{D}}\ell_{0}^{D-4}\Big(\ell_{0}^2-\frac{\alpha}{2}(D-3) \big(D(D-3)+4\big)  \Big)\ ,
	\cr
	&=&\frac{[(D-2)\text{!!}]^{2}(-k)^{\frac{D-1}{2}}}{(D-1)!\ell_{\rm eff}}\frac{\Omega_{D-2,k}}{8\pi G_{D}}\ell_{\rm eff}^{D-4}\left(\ell_{\rm eff}^{2}-2\alpha(D-3)(D-2)\right),
	\label{a1}
	\eea
	in which $\ell_{\rm eff}=\ell_{0}-\frac{\alpha(D-4)(D-3)}{2\ell_{0}}$ is the effective AdS radius.
	
	On the other hand, if one evaluates the on-shell
	Euclidean action of asymptotically AdS black holes directly on the uncorrected solution, the results will be incorrect as shown in the appendix \ref{AppendixA}. The discrepancy \eqref{dis} takes the form given in \eqref{Idiff} as can be seen easily as follows. We consider the difference of Euclidean action for two configurations $g_0+\delta_1g$ and $g_0$, where $g_0$ satisfies the leading order field equations. It follows that the variation of the bulk action simply yields a total derivative term
	\be
	\delta I_{\rm bulk}=-\frac{1}{16\pi G_D}\int d^{D}x\sqrt{g}(E_{\mu\nu}^{\text{bulk}}(g_0)\delta g^{\mu\nu}+\nabla_{\mu}\Theta^{\mu})\,.
	\ee
	Utilizing equation above and collecting terms up to first order in $\a$, one finds \cite{Reall:2019sah}
	\bea
	\left(I_{E}[g_0+\delta_1 g]-I_{E}[g_0]\right)|_\a&=&-\frac{1}{16\pi G_{D}}
	\int_{\partial\mathcal{M}|_{r={\infty}}}d^{D-1}x\sqrt{h} \left(K_{ab}-Kh_{ab}-E_{ab}^{\text{ct}}(h)\right)\delta_1 g^{ab}
	\nn\\
	&&+I_{\rm bulk}[g_0]|_{r=r_0+\Delta r}^{r=r_0}
	-\frac{1}{16\pi G_{D}}
	\int_{\partial\mathcal{M}|_{r=r_0}}d^{D-1}x\sqrt{h}n_\mu \Theta^\mu\,,
	\label{varI}
	\eea
	where $E_{ab}^{\text{ct}}(h)$ comes from the variation of the counterterm action w.r.t. $h_{ab}$. Thus the integrand in the right hand side of the first equality is proportional to $\langle T_{ab}\rangle\delta_1g^{ab}$ where $\langle T_{ab}\rangle$ is the holographic stress tensor \cite{Balasubramanian:1999re,deHaro:2000vlm}. In evaluating \eqref{varI}, we have performed the reparametrization of $r_0$ according to \eqref{dr0} and renamed $\tr_0$ back to $r_0$ so that the two on-shell actions (\ref{IBR},\ref{I0S}) are compared at the same temperature for a given value of $r_0$.  When $\delta_1 g$ is a perturbative solution about $g_0$, we find that the second line in \eqref{varI} vanishes and the difference between the two Euclidean actions solely comes from the contribution at asymptotic infinity.
	
	 As suggested in \cite[footnote 6]{Reall:2019sah}, one way to make the integral \eqref{varI}  vanish is by fixing the metric on the conformal boundary. This can be achieved by a proper rescaling of the time coordinate. For the corrected black hole solutions \eqref{EGBsol} in EGB model, the induced metric on the AdS boundary at $r\rightarrow\infty$  approaches
\be
ds_{D,\partial \mathcal{M}|_{r=\infty}}^{2}\approx-\frac{r^{2}}{\ell_{\text{eff}}^{2}}dt^{2}
+r^{2}d\Omega_{D-2,k}^{2}\,,
\ee
the effective AdS radius $\ell_{\text{eff}}$ is given by
\be
\frac{\ell_{\text{eff}}}{\ell_{0}}=1-\frac{\alpha(D-4)(D-3)}{2\ell_{0}^{2}}\,.
\ee
In the previous computation, the $\alpha$ dependent term in $\ell_{\rm eff}$ is
included in $\delta_1 g_{ab}$ and thus $\delta_1 g_{ab}\neq 0$. To fix the induced
metric on the conformal boundary,  we  can perform the following rescaling of the time coordinate
\be
t\rightarrow \gamma t\,,\qquad \gamma=\frac{\ell_{\text{eff}}}{\ell_{0}}\ ,
\ee
Accordingly, the thermodynamic quantities transform as follows
\be
T'\rightarrow \gamma^{-1}T'\,,\qquad   F'\rightarrow \gamma^{-1} F'\ ,
\ee
where $T'$ and $ F'$ denote the temperature and Helmholtz free energy calculated directly by using
the uncorrected solution in Appendix \ref{AppendixA}.   Again the modification of temperature due to the scaling can be absorbed into redefinition of $r_0'$ as
\be
r_0'=r_0+\delta r_0,\quad \delta r_{0}=\frac{\alpha(D-4)(D-3)r_{0}\left((D-3)k\ell_{0}^{2}
		+(D-1)r_{0}^{2}\right)}{2(D-3)k\ell_{0}^{4}-2(D-1)r_{0}^{2}\ell_{0}^{2}}\,,
\ee
in terms of which we find the Euclidean action is given by
\bea
T^{\rm RS}I_{E}^{\text{RS}}&=&F^{\rm RS}+{\rm mod}[D,2]\times a_1\,,\qquad T^{\rm RS}=\gamma^{-1}T'=\frac{D-3}{4 \pi  r_{0}}\big(k+\frac{(D-1) r_{0}^2}{(D-3)\ell_{0}^2}\big)\,,
\nn\\
F^{\rm RS}&=& \gamma^{-1} F'=\frac{r_{0}^{D-3}\left(k\ell_{0}^{2}-r_{0}^{2}\right) \omega_{D-2,k}}{16\pi\ell_{0}^{2}G_{D}}\nn\\
		&&\ -\alpha\frac{(D-3)(D-2)kr_{0}^{D-5} \left((D-2)k\ell_{0}^{2}+2(D-1)r_{0}^{2}\right) \omega_{D-2,k}}{16\pi\ell_{0}^{2}G_{D}}\,. \label{GBIRS}
\eea
To compare the results above with \eqref{IBR}, we set the parameters $\tr_0 =r_0$ so that these two Euclidean actions are defined at the same temperature. We then find that the result from the ordinary method agrees with the one from old RS method, namely
\be
I_{E}=I_{E}^{\text{RS}}\,.
\ee

From the steps carried out above, it is clear that to compute the first order corrections to the Euclidean action of static AdS black holes using the original RS method, one needs to first solve for the modified AdS radius. The complete structure of Gibbons-Hawking-York (GHY) terms and counterterms associated with the bulk higher derivative interactions is also indispensable to applying this approach. It can be seen from \eqref{EGBlag} and \eqref{Lct} that these terms are already complicated for the GB combination, let alone a generic higher-derivative bulk term. It is thus of great advantage to improve the RS method without having to deal with these terms, as in the case for the asymptotically-flat black holes. One may try to avoid constructing the higher derivative surface terms by adopting the background subtraction scheme \cite{Xiao:2023two} which is known to be ambiguous for stationary rotating black holes and may lead to incorrect result.\footnote{The method of subtracting the vacuum background can be also problematic for topological AdS black holes with hyperbolic horizon since the vacuum itself is already thermal with nonvanishing Euclidean action.}  These surface terms including both GHY terms and holographic counterterms were better studied for the Lovelock gravities, but definitely not for general higher-derivative curvature invariants. One can thus ask the question, can we improve the RS method for AdS black holes without having to construct the surface terms associated with higher derivative bulk interactions, as in the case of asymptotic flat black holes? As we will see in section 4, this can indeed be achieved.

	\section{Field redefinitions and thermodynamic equivalence}
	\label{sec:red}

	In the previous work \cite{Ma:2023qqj}, based on the RS method, we proved that, for asymptotically-flat black holes, the thermodynamic variables are manifestly invariant under field redefinitions. In the asymptotically-AdS case,  it becomes subtle to show that thermodynamic variables of AdS black holes are invariant under field redefinitions.  In this section, we will use examples of pure gravity and Einstein-Maxwell theory with generic 4-derivative corrections to show how to interpret the invariance of thermodynamic variables under field redefinitions.

	\subsection{Pure gravity with 4-derivative corrections}
	In the pure gravity case, we consider Einstein gravity with a negative cosmological constant extended by general 4-derivative couplings
	\be
	S_{2\partial+4\partial}=\frac{\sigma}{16\pi}\int d^Dx\sqrt{-g}( R+\ft{(D-1)(D-2)}{\ell_{0}^2}+ \sigma c_1 R^2 + \sigma c_2 R^{\mu\nu} R_{\mu\nu} + \sigma c_3 R^{\mu\nu\rho\sigma} R_{\mu\nu\rho\sigma})\,,
	\ee
	where for later, we use $\sigma$ to denote $1/G_D$. This theory admits a static AdS black hole solution. Up to first order in $c_i$, it is given by
	\bea
	ds_D^2 &=& - h(r) dt^2 + \fft{dr^2}{f(r)} + r^2 d\Omega_{D-2,k}^2\,,
	\nn\\
	h&=&\frac{r^2}{\ell_{0}^2}+k-\frac{2\mu}{r^{D-3}}+\Delta h\,,\qquad f=\frac{r^2}{\ell_{0}^2}+k-\frac{2\mu}{r^{D-3}}+\Delta f\,,\qquad \mu=\frac{r_{0}^{D-3} }{2 }\Big(k+\frac{r_{0}^2}{\ell_{0}^2}\Big)
	\label{EH-AdS}\,,
	\nn\\
	\Delta h&=&\Delta f=\frac{4 (D-4) (D-3) \mu ^2}{r^{2 (D-2)}}\sigma c_3
	+\frac{(D-4)   r^2}{(D-2) \ell_{0}^4}\sigma\Big( (D-1) Dc_1+ (D-1)c_2+2 c_3\Big)\,,
	\label{perturbed solution 2}
	\eea
	from which one clearly sees that the perturbations $\Delta h$ and $\Delta f$ do not modify the constant term $k$. The thermodynamic variables are
	\bea
	&&M=\frac{(  D-2)\mu  \omega_{D-2,k}}{8\pi}\sigma\Big(1-\frac{2\sigma}{\ell_{0}^2}\big( (D-1) \left( Dc_1+c_2\right)-2 (D-4)c_3\big)\Big)\,,
	\cr
	&&S=\frac{  \omega_{D-2,k}}{4}\sigma r_{0}^{D-2}+\Delta S\,, \cr
	&&\Delta S=-\frac{\sigma ^2 r_{0}^{D-4} \omega _{D-2,k}}{4 \ell_{0}^2 ((D-1) r_{0}^2+(D-3) k \ell_{0}^2)}
	\Big[
	(D-1) (Dc_1 +c_2) r_{0}^2 (3 (D-2) r_{0}^2+2 (D-3) k \ell_{0}^2)\cr
	&&-c_3 \big(2 (D-3) (D^2-2 D-2) k \ell_{0}^2 r_{0}^2+(D-4) (D-2) (D+3) r_{0}^4+(D-3) (D-2)^2 k^2 \ell_{0}^4\big)
	\Big]\,,
	\cr
	&&T=\frac{D-3}{4 \pi  r_{0}}\Big(k+\frac{(D-1) r_{0}^2}{(D-3) \ell_{0}^2}\Big)+\Delta T\,,\cr
	&&\Delta T=\frac{(D-4) \sigma }{4 \pi  (D-2) \ell_{0}^4 r_{0}^3 \left((D-1) r_{0}^2+(D-3) k \ell_{0}^2\right)}\cr
	&&\times\Big[(D-1) \left( Dc_1+c_2\right) r_{0}^4 \left((D-2) (D-1) r_{0}^2+(D-3) D k \ell_{0}^2\right)
	-c_3 \big((D-3)^2 (D-2)^2 k^3 \ell_{0}^6
	\cr
	&&+(D-3) (D-2) (3 D^2-11 D+12) k^2 \ell_{0}^4 r_{0}^2+(D-3) (3 D^3-13 D^2+18 D-12) k \ell_{0}^2 r_{0}^4\cr
	&&+(D-2) (D-1)(D^2-3 D-2) r_{0}^6\big)
	\Big]\,.
	\label{corrected pure gravity}
	\eea

	We now investigate how these thermodynamic variables transform under the field redefinitions below
	\be
	g_{\mu\nu}	\rightarrow g'_{\mu\nu}=g{}_{\mu\nu}+\lambda_{0}g{}_{\mu\nu}+\sigma\lambda_{1}
	R{}_{\mu\nu}+\sigma\lambda_{2}g_{\mu\nu}R\,,
	\label{redefg}
	\ee
	which should be compensated by transformations of the parameters in the action as
	\bea
	\sigma	&\rightarrow & \sigma'=\sigma-\frac{D-2}{2}\sigma\Big(\lambda_{0}+\sigma\frac{(D-1)(\lambda_{1}
		+D\lambda_{2})}{\ell_{0}^{2}}\Big)\,,
	\nn\\
	\ell_{0}& \rightarrow & \ell_{0}'=\ell_{0}+\frac{\ell_{0}}{2}\Big(\lambda_{0}-\sigma\frac{(D-1)(D-2)
		(\lambda_{1}+D\lambda_{2})}{2\ell_{0}^{2}}\Big)\,,
	\nn\\
	c_{1}&\rightarrow & c_{1}'=c_{1}-(\frac{1}{2}\lambda_{1}+\frac{D-2}{2}\lambda_{2}),\quad c_{2}	\rightarrow  c_{2}'=c_{2}+\lambda_{1},\quad c_{3} \rightarrow  c_{3}'=c_{3}\,,
	\label{field redefinition}
	\eea
	so that the actions before and after the redefinition satisfy
	\be
	I[g,\,c]=I'[g',\, c']\,.
	\label{redefS}
	\ee
	Starting from a solution $g_{\mu\nu}$ of the original theory, one immediately obtains the solution $g'_{\mu\nu}$ in the redefined theory from \eqref{redefg}, which is of the form below
	\bea
	ds_D^{'2} &=&- h'(r) dt^2 + \fft{dr^2}{f'(r)} + \rho^2(r) d\Omega_{D-2,k}^2\,,
	\nn\\
	\rho^2(r)&=&r^2\left[1+\Big(\lambda _0-\frac{(D-1) \sigma ( \lambda _1+D\lambda _2)}{\ell_{0}^2}\Big)\right].
	\label{newsol}
	\eea
	In order to reexpress the solution above \eqref{newsol} into the standard form \eqref{perturbed solution 2}, we choose $\rho$ to be the new radial coordinate and a new time $t'$ related to $t$ by
	\be
	t'=t\left[1+\frac12\Big(\lambda _0-\frac{(D-1) \sigma ( \lambda _1+D\lambda _2)}{\ell_{0}^2}\Big)\right].
	\ee
	In terms of $\rho,\,t'$, the new metric takes the standard form
	\be
	ds_D^{'2} =- (h_0(\rho)+\Delta h'(\rho)) dt^{'2} + \frac{d{\rho}^2}{f_0(\rho)+\Delta f'(\rho)} + \rho^2 d\Omega_{D-2,k}^2\,,
	\label{newsol1}
	\ee
	in which the corrections $\Delta h'(\rho)$ and $\Delta f'(\rho)$ do not contain a constant piece as the $k$ term appearing in $ h_0(\rho)$ and $ f_0(\rho)$. The relation between $t'$ and $t$ implies that the energies of black hole solutions before and after the field redefinitions are related by
	\be
	M'=M\left[1-\frac12\Big(\lambda _0-\frac{(D-1) \sigma ( \lambda _1+D\lambda _2)}{\ell_{0}^2}\Big)\right],
	\ee
	which is also confirmed by an explicit computation.
	Clearly, for the mass to be invariant under field redefinitions, we have to choose
	\be
	\lambda _0=\frac{(D-1) \sigma ( \lambda _1+D\lambda _2)}{\ell_{0}^2}\,.
	\label{l0}
	\ee
	 This choice of $\l_0$ can also be seen from a different perspective. Under the field redefinition \eqref{redefg}, assuming the period of imaginary time remains the same, the difference of on-shell Euclidean action is given by an expression similar to \eqref{Idiff}
		\be
		\delta I_E\propto \int_{\partial\mathcal{M}|_{r={\infty}}}d^{D-1}x\sqrt{h}T^{(0)ab}\delta g_{ab},\quad  \delta g_{\m\m}=g'_{\m\n}-g_{\m\n}\ .
		\ee
		For asymptotically AdS black holes,  due to the fact that near the AdS boundary
		\be
		\delta g_{\m\n}\approx \widetilde{d}\, g_{(0)\m\n},\quad \widetilde{d}=\Big(\lambda _0-\frac{(D-1) \sigma ( \lambda _1+D\lambda _2)}{\ell_{0}^2}\Big)\ ,
		\ee
		it appears that $\delta I_E$ vanishes because $T^{(0)ab}g_{(0)ab}=0$ for AdS black hole with boundary topology $\mathbb{R}\times S^{D-2}$. Thus one may have the illusion that for any field redefinition leaves invariant the black hole thermodynamics. However, this is not the case because the field redefinition has modified the induced metric on the conformal boundary which is given by
		\be
		ds_{D,\partial \mathcal{M}|_{r=\infty}}^{'2}\approx (1+\widetilde{d})\left(-\frac{r^{2}}{\ell_{\text{eff}}^{2}}dt^{2}+r^{2}d\Omega_{D-2,k}^{2}\right)\ ,
		\ee
		To preserve the induced metric under field redefinition, we need to set $\widetilde{d}=0$ which
		also leads to the specific choice of $\l_0$ as in \eqref{l0}.
	
	When \eqref{l0} is satisfied, we also find that  the temperature computed from $g'_{\m\n}$ is equal to the one computed using $g_{\m\n}$, namely $T'=T$. Since \eqref{redefS} already implies the equality between the on-shell Euclidean actions \footnote{For the on-shell action to be finite, one must supplement the bulk action with Gibbons-Hawking term and counterterm on the AdS boundary. For general curvature squared gravity, one method of deriving these surface terms was proposed in \cite{Cremonini:2009ih}.}
	\be
	I_E[T]=I'_E[T']\,.
	\ee
	The equivalence between $T$ and $T'$ suggests that $I_E$ and $I'_E$ have the same form as a function of temperature. Therefore, we show that field redefinitions obeying \eqref{l0} do not affect the form of the on-shell action as function of the temperature, which implies the invariance of other thermodynamic variables.
	
	On the other hand, one can also carry out redefinitions of parameters \eqref{field redefinition} directly on the solution \eqref{perturbed solution 2} and evaluate the corresponding thermodynamic variables using the new solution. We notice that, if one makes coordinates transformation $r\rightarrow\r,\,t\rightarrow t'$ and
	shift the parameter $\m$ according to
	\be
	\m\rightarrow\mu'=\mu+\frac{(D-3) \mu}{2} \Big(\lambda _0-\frac{(D-1) \sigma  }{\ell_{0}^2}\left(\lambda _1+D \lambda _2\right)\Big)\,,
	\label{parameter redefinition1}
	\ee
	then resulting solution becomes identical to the one in \eqref{newsol1}. As the solution agrees, the corresponding thermodynamic variables must be identical.
	We thus learn that in a higher derivative theory of gravity, matching physical quantities before and after the field redefinition may require shifting the integration constants in an appropriate way. This fact allows physical quantities to depend on parameters in the Lagrangian that are not inert under field redefinitions, because their effects can be canceled by transformations of the integration constants.
	
	In the asymptotically Minkowski case corresponding to $\ell_{0}\rightarrow\infty$, we usually do not consider rescaling of the metric. Upon setting $\l_0=0$, we see that the shift of $\m$ vanishes, consistent with the findings of \cite{Ma:2023qqj} where it was shown that the thermodynamic variables of asymptotically flat black hole solutions are manifestly invariant under  field redefinitions since parameters affected by field redefinitions do not enter the expressions of thermodynamic variables.

	\subsection{Einstein-Maxwell theory extended by general 4-derivative terms}
	
	We begin with the $D$-dimensional Einstein-Maxwell theory of the form
	\bea
	S_{EM}&=&\frac{1}{16\pi }\int d^Dx\sqrt{-g}\Big(\sigma R-\frac{1}{4g^2}F^2+\frac{(D-1)(D-2)}{\ell_{0}^2}\sigma\Big),
	\label{DEM}
	\eea
	which admits a static charged AdS black hole
	\bea
	ds_D^2 &=&- h_0(r) dt^2 + \fft{dr^2}{f_0(r)} + r^2 d\Omega_{D-2,k}^2\,,\qquad A_{(1)}=\psi_0(r) dt\,,
	\cr
	h_0&=&f_0=\frac{r^2}{\ell_{0}^2}+k-\frac{2\mu}{r^{D-3}}+\frac{q^2}{r^{2(D-3)}}\,,
	\qquad \psi_{0}=\sqrt{\frac{2(D-2)\sigma}{D-3}}\frac{gq}{r^{D-3}}\,.\label{EM-AdS}
	\eea
	Various thermodynamic variables are given by
	\bea
	M_0&=&\frac{\left(  D-2\right)  \omega_{D-2,k}}{8\pi}\,\sigma\mu\,,\qquad T_0=\frac{D-3}{4 \pi  r_{0}}\Big(k-\frac{q^2}{r_{0}^{2 (D-3)}}+\frac{(D-1) r_{0}^2}{(D-3) \ell_{0}^2}\Big),
	\cr
	S_0&=&\frac{  \omega_{D-2,k}}{4}\sigma r_{0}^{D-2}\,,\qquad \Phi_{e,0}=A_t|^{\infty}_{r_0}=-\sqrt{\frac{2(D-2)\sigma}{D-3}}\frac{gq}{r_0^{D-3}}
	\,,
	\cr
	Q_{e,0}&=&\frac{1}{16 \pi g^2}\int \star F =-\frac
	{\sqrt{2\sigma\left(  D-2\right)  \left( D-3\right)  }\omega_{D-2,k}}{16\pi g }q\,,
	\eea
	where $r_0$ denotes the location of the horizon and obeys
	\be
	\mu=\frac{q^2}{2 r_{0}^{D-3}}+\frac{r_{0}^{D-3} }{2 }(k+\frac{r_{0}^2}{\ell_{0}^2})
	\,.
	\ee
	The following parity even 4-derivative terms can be added to \eqref{DEM}
	\bea
	\mathcal{L}_{4\partial} &=&  \sigma^2c_1 R^2 +\sigma^2 c_2 R^{\mu\nu} R_{\mu\nu} +\sigma^2 c_3 R^{\mu\nu\rho\sigma} R_{\mu\nu\rho\sigma}+  \frac{\sigma c_4}{g^2} R F^2 + \frac{\sigma c_5}{g^2} R^{\mu\nu} F_{\mu\rho} F_\nu{}^\rho\cr
	&&\qquad +   \frac{\sigma c_6}{g^2} R^{\mu\nu\rho\sigma} F_{\mu\nu} F_{\rho\sigma}+  \frac{c_7}{g^4} (F^2)^2 +  \frac{c_8}{g^4} F^{\mu}{}_\nu F^\nu{}_\rho F^\rho{}_\sigma F^{\sigma}{}_\mu \,,
	\label{EMAction}
	\eea
	which modifies the static charged AdS black hole solutions as follows
	\be
	h= h_0 + \Delta h\,,\qquad f= f_0 + \Delta f\,,\qquad \psi= \psi_0 + \Delta \psi\,.
	\label{perturbed solution3}
	\ee
	The perturbations about the leading order solution are quite complicated. Thus we postpone their presentations in Appendix \ref{Perturbed solution}. The mass and electric charge of the corrected solution are given by
	\be
	M=M_0\Big[1-\frac{2\sigma}{\ell_{0}^2}\big( (D-1) \left( Dc_1+c_2\right)-2 (D-4)c_3\big)\Big],\qquad Q_e=Q_{e,0}\,,
	\label{mq}
	\ee
	The corrected entropy, temperature and electric potential are a bit lengthy. Their expressions are listed in  Appendix \ref{Corrected Thermodynamics}.  Different from the asymptotic Minkowski case \cite{Ma:2023qqj}, the thermodynamic variables of AdS black holes depend on parameters which transform under field redefinitions. Thus it is natural to understand how these physical quantities remain invariant under field redefinitions. Consider the general field redefinitions below
	\bea
	g_{\mu\nu}&\rightarrow& g'_{\mu\nu}=g_{\mu\nu}+\lambda_0 g_{\mu\nu}+\sigma\lambda_1 R_{\mu\nu}+\sigma\lambda_2g_{\mu\nu}R+\frac{\lambda_3}{g^2}F_{\mu\rho}F_{\nu}^{\ \rho}+\frac{\lambda_4}{g^2}g_{\mu\nu}F^2\,,
	\cr
	A_{\m}&\rightarrow&A_{\m}'=A_{\m}+\lambda_5A_{\m}\,,
	\label{field redefinition 1}
	\eea
	to ensure that the actions before and after the redefinition satisfy
	\be
	I[g_{\m\n},\,A_{\m},\,c]=I'[g_{\m\n}',\,A_\m',\,c']\,,
	\label{redefS2}
	\ee
	the parameters in the action must be shifted according to
	\begin{align}
		& \sigma\rightarrow\sigma'=\sigma-\frac{D-2}{2}\Big(\lambda_{0}
		+\frac{(D-1)(\lambda_{1}+D\lambda_{2})}{\ell_{0}^{2}}\sigma\Big)\sigma\,,
		\nonumber \\
		& \ell_{0}\rightarrow\ell_{0}'=\ell_{0}
		+\frac{\ell_{0}}{2}\Big(\lambda_{0}-\sigma\frac{(D-1)(D-2)(\lambda_{1}
			+D\lambda_{2})}{2\ell_{0}^{2}}\Big),
		\nonumber \\
		& \frac{1}{g^{2}}\rightarrow\frac{1}{g'^{2}}=\frac{1}{g^{2}}
		-\frac{1}{2g^{2}}\Big((D-4)\lambda_{0}+4\lambda_{5}
		-\frac{4(D-2)(D-1)(\lambda_{3}+D\lambda_{4})}{\ell_{0}^{2}}\sigma\Big),
		\nonumber \\
		& c_{1}\rightarrow c_{1}'=c_{1}-\frac{1}{2}\lambda_{1}-\frac{D-2}{2}\lambda_{2},\quad c_{2}\rightarrow c_{2}'=c_{2}+\lambda_{1},\quad c_{3}\rightarrow c_{3}'=c_{3}\,,
		\nonumber \\
		& c_{4}\rightarrow c_{4}'=c_{4}+\frac{1}{8}\lambda_{1}+\frac{D-4}{8}\lambda_{2}
		-\frac{1}{2}\lambda_{3}-\frac{D-2}{2}\lambda_{4}\,,
		\nonumber \\
		& c_{5}\rightarrow c_{5}'=c_{5}-\frac{1}{2}\lambda_{1}+\lambda_{3},\quad c_{6}\rightarrow c_{6}'=c_{6}\,,
		\nonumber\\
		& c_{7}\rightarrow c_{7}'=c_{7}+\frac{1}{8}\lambda_{3}+\frac{D-4}{8}\lambda_{4},\quad c_{8}\rightarrow c_{8}'=c_{8}-\frac{1}{2}\lambda_{3}\,.
		\label{field redefinition 2}
	\end{align}
	Given a solution $g_{\m\n},\,A_{\m}$ of the original theory, one immediately obtains a solution in the redefined theory using \eqref{field redefinition 1}, which is of the form
	\bea
	ds_D^{'2} &=&- h'(r) dt^2 + \fft{dr^2}{f'(r)} + \rho^2(r) d\Omega_{D-2,k}^2\,,\qquad A'_{(1)}=(1+\l_5)\psi(r) dt\,,\label{newsol2}
	\\
	\rho^2(r)&=&r^2\left[1+\Big(\lambda _0-\frac{(D-1) \sigma ( \lambda _1+D\lambda _2)}{\ell_{0}^2}\Big)+\frac{(D-3) \sigma  q^2}{r^{2 (D-2)}}\big(\lambda _1-(D-4) \lambda _2-4 (D-2) \lambda _4
	\big)\right], \nn
	\eea
	where the explicit forms of $\Delta h$ and $\Delta f$ are not important for later discussions.
	In order to reexpress the metric above \eqref{newsol2} into the standard form \eqref{perturbed solution3} where the constant term in $h'$ and $f'$ remains to be $k$, we choose $\rho$ to be the new radial coordinate and a new time $t'$ related to $t$ by
	\be
	t'=t\left[1+\frac12\Big(\lambda _0-\frac{(D-1) \sigma ( \lambda _1+D\lambda _2)}{\ell_{0}^2}\Big)\right].
	\ee
	Similar to the pure gravity case, the relation between $t'$ and $t$ implies that the energies of black hole solutions before and after the field redefinitions are related by
	\be
	M'=M\left[1-\frac12\Big(\lambda _0-\frac{(D-1) \sigma ( \lambda _1+D\lambda _2)}{\ell_{0}^2}\Big)\right].
	\ee
	Thus for the mass to be invariant under field redefinitions, we must choose
	\be
	\lambda _0=\frac{(D-1) \sigma ( \lambda _1+D\lambda _2)}{\ell_{0}^2}\,.
	\label{l02}
	\ee
	With this choice, we find the temperature and the electric potential of the solution in the redefined theory take the form
	\be
	T'=T,\quad \Phi_e'=(1+\lambda_5)\Phi_{e}\,,
	\label{thermodynamic method 2}
	\ee
	thus field redefinitions with $\lambda_5=0$ preserve the electric potential.
	We recall that \eqref{redefS2} already means  on-shell actions associated with solutions in the original theory and redefined theory are identical
	\be
	I_E(T,\Phi_e)=I'_E(T',\,\Phi_e')\,.
	\ee
	Thus $T=T'$ and $\Phi_e=\Phi_e'$ indicate that $I'_E$ has the same the form as $I_E$. For static charged AdS black hole solutions, the equivalence of the on-shell actions leads to equivalence of the Gibbs free energies and thus all the other thermodynamic variables.

	On the other hand, one can also carry out the redefinition of the parameters \eqref{field redefinition 2} directly
	on the solution \eqref{perturbed solution3}. The resulting solution matches with the one obtained from \eqref{field redefinition 1} after performing the coordinates transformation $r\rightarrow \rho$, $t\rightarrow t'$ as well as the shift of the integration constants below
	\bea
	q&\rightarrow& q+\frac{q}{4}\left[
	2 (D-3) \lambda _0+\frac{(D-2) (D-1) \sigma }{\ell_{0}^2}\big(\lambda _1+D \lambda _2-4 \left(\lambda _3+D \lambda _4\right)\big)
	\right],
	\cr
	\mu&\rightarrow& \mu+\frac{(D-3) \mu}{2} \Big(\lambda _0-\frac{(D-1) \sigma  }{\ell_{0}^2}(\lambda _1+D \lambda _2)\Big)\,.
	\label{parameter redefinition}
	\eea
	Therefore in the case of charged AdS black holes, we see again that  when the physical quantities are expressed in terms of integration constants of the solution, matching results before and after the field redefinition may require shifting the integration constants in an appropriate way. Therefore physical quantities are allowed to depend on parameters that vary under field redefinitions whose effects can be canceled by transformations of the integration constants.
	
	In the asymptotically Minkowski case corresponding to $\ell_{0}\rightarrow\infty$, upon setting $\l_0=0$, we see that the shift of integration constants vanishes, consistent with the findings of \cite{Ma:2023qqj} where it was shown that the thermodynamic variables of asymptotically flat static charged black holes are thus manifestly invariant without transforming the integration constants, as they are independent of parameters affected by field redefinitions.

	\section{Improved Reall-Santos method for AdS black holes }
	\label{sec:improved}
	
	Having understood under which circumstance, field redefinitions do not modify the thermodynamic variables, one is allowed to choose any equivalent field frame where the computation is to be carried out. We recall that for AdS black holes in models with general 4-derivative terms, without applying the rescaling of the time coordinate,
	the perturbative solution $\delta_1g_{\m\n}$ does not fall off fast enough near the AdS boundary. To remedy this issue, one can choose a particular field frame in which perturbations fall off sufficiently fast. It turns out that this particular frame has another advantage that higher derivative surface terms are not needed for the implementation of variational principle and removal of divergences. Below, we will show that Weyl tensor squared has this property. By explicit computations, we show the equivalence between results obtained from the ordinary method and the improved RS method, meaning that one can obtain all the interesting physical observables in Einstein-Maxwell-Weyl gravity without actually solving the perturbations caused by 4-derivative terms. This result can be further used to obtain thermodynamic variables for AdS black holes in general 4-derivative extensions of Einstein-Maxwell theory which is related to Einstein-Maxwell-Weyl by appropriate field redefinitions.

	\subsection{Thermodynamics of AdS black holes in  Einstein-Maxwell-Weyl gravity}
	For asymptotically AdS black holes, the Weyl tensor squared term is rather special. It's finite on the solution without the need of adding boundary counterterms which is consistent with the vanishing of potential boundary terms arising in the variational principle. Thus, for Einstein-Maxwell-Weyl gravity, we only need to include the GHY term and standard surface counterterms for the 2-derivative theory. The action of Einstein-Maxwell-Weyl gravity is given by
	\bea
	I_{EMW}&=&\frac{1}{16\pi}\int d^{D}x\sqrt{-g}(\mathcal{L}_{EM}+\mathcal{L}_{C^{2}})+I_{\rm GHY}-I_{\rm ct}\,,
	\label{WeylAction}
	\\
	\mathcal{L}_{EM}&=&\sigma\big(R+\frac{(D-1)(D-2)}{\ell_{0}^{2}}\big)
	-\frac{1}{4g^{2}}F^{2}\,,
	\cr
	\mathcal{L}_{C^{2}}&=&\sigma^{2}c_{3}C^{\mu\nu\rho\sigma}C_{\mu\nu\rho\sigma}+\frac{\sigma c_{6}}{g^{2}}C^{\mu\nu\rho\sigma}F_{\mu\nu}F_{\rho\sigma}+
	\frac{c_{7}}{g^{4}}(F^{2})^{2}+
	\frac{c_{8}}{g^{4}}F^{\mu}{}_{\nu}F^{\nu}{}_{\rho}F^{\rho}{}_{\sigma}
	F^{\sigma}{}_{\mu}\,,\nn
	\eea
	where the GHY boundary term and surface counterterms for $D$-dimensional Einstein theory are
	\bea
	I_{\mathrm{GHY}}&=&\frac{\sigma}{8\pi }\int d^{D-1}x\sqrt{-h}K\,,\ \ \ I_{\mathrm{ct}}=\frac{\sigma}{16\pi }\int d^{D-1}x\sqrt{-h}
	\mathcal{L}_{\mathrm{ct}}\,, \cr
	\mathcal{L}_{\mathrm{ct}}&=&\frac{2(D-2) }{\ell_{0}}
	+\frac{\ell_{0}}{D-3}  \mathcal{R}+\frac{\ell_{0}^3}{(D-5)(D-3)^2}  \big(\mathcal{R}_{ab}^2-\frac{D-1}{4(D-2)}\mathcal{R}^2\big)+\cdots\,.
	\eea
	Static charged AdS black hole solutions and their thermodynamic variables in the theory above can be obtained from the general results \eqref{perturbed solution3}  by choosing the coefficients below
	\bea
	c_1=\frac{2}{(D-1)(D-2)}c_3,\ \ c_2=-\frac{4}{D-2}c_3,\ \
	c_4=\frac{2}{(D-1)(D-2)}c_6,\ \  c_5=-\frac{4}{D-2}c_6\,.
	\eea
	To compare with the results obtained from improved RS method, we redefine the parameter $r_0,\,q$
	\bea
	r_0=\widetilde{r}_0+\delta \widetilde{r}_0\,,\qquad q=\widetilde{q}+\delta \widetilde{q}\,,
	\eea
	where $\delta \widetilde{r}_0$ is given by
	\begin{align}
		\delta\tilde{r}_{0} & =\frac{(D-3)\sigma}{(D-1)(3D-7)\ell_{0}^{2}\widetilde{r}_{0}^{2D+13}
			\big((D-1)^{2}\widetilde{r}_{0}^{4(D-2)}-(D-3)^{2}\ell_{0}^{4}
			(\widetilde{q}^{2}-k\widetilde{r}_{0}^{2(D-3)})^{2}\big)}\nonumber \\
		& \times\Bigg((D-3)\widetilde{q}^{4}\ell_{0}^{4}\widetilde{r}_{0}^{2(D+6)}
		\Big[8\big((D-4)(D-3)k\ell_{0}^{2}+(D-6)(D-1)\widetilde{r}_{0}^{2}\big)
		(D-2)(D-1)\nonumber \\
		& \times(2c_{7}+c_{8})+4(D-3)\big(2(D-4)(D-2)(2D-5)k\ell_{0}^{2}
		+(D-1)\left(4D^{2}-27D+40\right)\widetilde{r}_{0}^{2}\big)c_{6}\nonumber \\
		& +\big((D-4)(D-3)(D-2)(13D-31)k\ell_{0}^{2}+(D-1)(13D^{3}-123D^{2}
		+364D-344)\widetilde{r}_{0}^{2}\big)c_{3}\Big]\nonumber \\
		& -(3D-7)\widetilde{q}^{2}\ell_{0}^{2}\widetilde{r}_{0}^{4D+6}
		\Big[2(D-3)(D-1)\big(2(D^{2}-6D+6)k\widetilde{r}_{0}^{2}\ell_{0}^{2}
		+(D-4)(D-3)k^{2}\ell_{0}^{4}\nonumber \\
		& +(D-4)(D-1)\widetilde{r}_{0}^{4}\big)c_{6}+\big((3D^{3}-21D^{2}
		+52D-40)\widetilde{r}_{0}^{4}+(D-4)(D-3)(3D-8)k^{2}\ell_{0}^{4}\nonumber \\
		& +2(D-4)(D-3)(3D-4)k\widetilde{r}_{0}^{2}\ell_{0}^{2}\big)(D-1)c_{3}\Big]
		+(D-4)(D-1)(3D-7)\widetilde{r}_{0}^{6D}(k\ell_{0}^{2}+\widetilde{r}_{0}^{2})^{2}
		\nonumber \\
		& \times\left((D-3)(D-2)k\ell_{0}^{2}+(D-1)D\widetilde{r}_{0}^{2}\right)c_{3}
		-(D-4)(D-3)^{2}\widetilde{q}^{6}\widetilde{r}_{0}^{18}\ell_{0}^{6}
		\big((D-3)(7D-16)c_{3}\nonumber \\
		& +2(D-3)(5D-11)c_{6}+8(D-2)(D-1)(2c_{7}+c_{8})\big)\Bigg),
	\end{align}
	and $\delta \widetilde{q}$ is given by
	\begin{align}
		\delta\widetilde{q} & =\frac{(D-3)^{2}\sigma\widetilde{q}\widetilde{r}_{0}^{-2(D+4)}}{(D-1)(3D-7)
			\ell_{0}^{2}\big((D-3)\ell_{0}^{2}(\widetilde{q}^{2}
			-k\widetilde{r}_{0}^{2(D-3)})+(D-1)\widetilde{r}_{0}^{2D-4}\big)}\nonumber \\
		& \times\Bigg((D-3)^{2}\widetilde{q}^{4}\widetilde{r}_{0}^{12}
		\ell_{0}^{4}\big(2(D-3)(5D-11)c_{6}+8(D-2)(D-1)(2c_{7}+c_{8})\nonumber \\
		& +(D-3)(7D-16)c_{3}\big)+(3D^{2}-10D+7)\big((D-4)(D-1)
		(k\ell_{0}^{2}+\widetilde{r}_{0}^{2})c_{3}\nonumber \\
		& +2((D-3)k\ell_{0}^{2}+(1-D)\widetilde{r}_{0}^{2})c_{6}\big)(k\ell_{0}^{2}
		+\widetilde{r}_{0}^{2})\widetilde{r}_{0}^{4D}
		+2\widetilde{q}^{2}\ell_{0}^{2}\widetilde{r}_{0}^{2D+6}\Big[(2c_{7}+c_{8})\nonumber \\
		& \times8(D-2)(D-1)\left((D-1)\widetilde{r}_{0}^{2}
		-(D-3)k\ell_{0}^{2}\right)-\big((D-1)(3D^{2}-20D+30)\widetilde{r}_{0}^{2}\nonumber \\
		& +(3D^{3}-15D^{2}+10D+20)k\ell_{0}^{2}\big)(D-3)c_{3}-
		\big((D-1)(3D^{2}-23D+36)\widetilde{r}_{0}^{2}\nonumber \\
		& +(3D^{3}-12D^{2}-5D-38)k\ell_{0}^{2}\big)(D-3)c_{6}\Big]\Bigg),
	\end{align}
	so that in terms of the new parameter $\widetilde{r}_0, \widetilde{q}$, the temperature and electric potential does not have explicit $c_i$ dependence
	\bea
	T=\frac{D-3}{4\pi\widetilde{r}_{0}}\Big(k-
	\frac{\widetilde{q}^{2}}{\widetilde{r}_{0}^{2(D-3)}}
	+\frac{(D-1)\widetilde{r}_{0}^{2}}{(D-3)\ell_{0}^{2}}\Big),\qquad \Phi_{e}=-\sqrt{\frac{2(D-2)\sigma}{D-3}}
	\frac{g\widetilde{q}}{\widetilde{r}_{0}^{D-3}}\,.
	\eea
Accordingly, the Gibbs free energy is obtained as follows
	\bea
&&\!\!\!G_{EMW} (T,\Phi_{e},c_{i})=M-TS-\Phi_{e}Q_{e}=-
	\frac{\sigma\omega_{D-2,k}}{16\pi\ell_{0}^{2}\widetilde{r}_{0}^{D-3}}
	\left(\widetilde{r}_{0}^{2(D-3)}(\widetilde{r}_{0}^{2}-k\ell_{0}^{2})
	+\widetilde{q}^{2}\ell_{0}^{2}\right)+\Delta G_{EMW}\,,\cr
&&\,\,	\Delta G_{EMW}  = \frac{(D-3)(D-2)\sigma^{2}\omega_{D-2,k}}{16\pi(D-1)(3D-7)\ell_{0}^{4}
		\widetilde{r}_{0}^{3D-7}}\Big[-(D-2)(D-1)(3D-7)
	\widetilde{r}_{0}^{4(D-3)}(k\ell_{0}^{2}+\widetilde{r}_{0}^{2})^{2}c_{3}
	\cr
\,\,&& +2(D-3)(D-1)(3D-7)\widetilde{q}^{2}\ell_{0}^{2}\widetilde{r}_{0}^{2(D-3)}
	(k\ell_{0}^{2}+\widetilde{r}_{0}^{2})(c_{3}+c_{6})-\big((D-3)(7D-16)c_{3}
	\cr
\,\,&& +2(D-3)(5D-11)c_{6}+8(D-2)(D-1)(2c_{7}+c_{8})\big)(D-3)\widetilde{q}^{4}
	\ell_{0}^{4}\Big].
	\eea
	Substituting the solution with redefined parameter $\widetilde{r}_0, \widetilde{q}$ into the action \eqref{WeylAction}, we find that the Euclidean action $I_{EMW}$ satisfies
	\bea
	T I_{EMW}=G_{EMW}+\text{mod}[D,2]\times a_{EMW}\,, \label{WeylI}
	\eea
	where $a_{EMW}$ appears in odd spacetime dimensions and takes the form
	\bea
	a_{EMW}=\frac{2(-k)^{\frac{1}{2}(D-1)}[(D-2)!!]^2}{(D-1)!}\frac{a^*}{\ell_{0}}\,,\qquad   a^*=\frac{\sigma\Omega_{D-2,k}}{16\pi}\ell_{0}^{D-2}\,,
	\eea
	Thus the Weyl tensor squared does not contribute to the central
	charge. Notice that in Einstein-Weyl gravity, the effective AdS radius $\ell_{\rm eff}=\ell_{0}$.
	
	On the other hand, if adopting the improved RS method,  up to order $c_i$,  the on-shell Euclidean action of AdS black hole in Einstein-Maxwell-Weyl theory could be obtained by evaluating the total action \eqref{WeylAction} on the uncorrected solution. The result is given by
	\be
	T I_{EMW}^{\rm RS}(T,\Phi_{e},c_{i})=G_{EMW}^{\rm RS}+\text{mod}[D,2]
	\times a_{EMW}\,,\label{Weyl 2 RS}
	\ee
	where
	\bea
	&&T=\frac{D-3}{4\pi r_{0}}\Big(k-\frac{q^{2}}{r_{0}^{2(D-3)}}+\frac{(D-1)r_{0}^{2}}{(D-3)\ell_{0}^{2}}\Big), \qquad \Phi_{e}=-\sqrt{\frac{2(D-2)\sigma}{D-3}}\frac{gq}{r_{0}^{D-3}}\,,
	\cr
	&&G_{EMW}^{\rm{RS}}  (T,\Phi_{e},c_{i})=-\frac{\sigma\omega_{D-2,k}}{16\pi\ell_{0}^{2}r_{0}^{D-3}}
	\big(r_{0}^{2(D-3)}(r_{0}^{2}-k\ell_{0}^{2})+q^{2}\ell_{0}^{2}\big)+\Delta G_{EMW}^{\rm RS}\,,\cr
	&&\Delta G_{EMW}^{\rm RS} =\frac{(D-3)(D-2)\sigma^{2}\omega_{D-2,k}}{16\pi(D-1)(3D-7)
		\ell_{0}^{4}r_{0}^{3D-7}}\Big[-(D-2)(D-1)(3D-7)r_{0}^{4(D-3)}(k\ell_{0}^{2}
	+r_{0}^{2})^{2}c_{3}\cr
	&& +2(D-3)(D-1)(3D-7)q^{2}\ell_{0}^{2}r_{0}^{2(D-3)}(k\ell_{0}^{2}
	+r_{0}^{2})(c_{3}+c_{6})-\Big((D-3)(7D-16)c_{3}
	\cr
	&& +2(D-3)(5D-11)c_{6}+8(D-2)(D-1)(2c_{7}+c_{8})\Big)(D-3)q^{4}\ell_{0}^{4}\Big].
	\eea
	To compare \eqref{WeylI} with \eqref{Weyl 2 RS}, we set $\widetilde{r}_0=r_0$, $\widetilde{q}=q$, so that these two Euclidean actions
	are defined at the same temperature and electric potential. We then find that
	the result from the ordinary method agrees with the one from improved RS method
	\be
	I_{EMW}=I_{EMW}^{\rm RS}\,,
	\ee
	which clearly states that the improved RS method is valid for computing thermodynamic variables in Einstein-Maxwell-Weyl theory  \eqref{WeylAction}.
	
	\subsection{Thermodynamics of AdS black holes in general 4-derivative gravities}
	
	In the previous section,  we have examined the validity of the improved RS method in the
	Einstein-Maxwell-Weyl theory. In section 3, we have shown that field redefinitions obeying
	\be
	\lambda _0=\frac{(D-1) \sigma ( \lambda _1+D\lambda _2)}{\ell_{0}^2}\,,\qquad
	\lambda_5=0\,,
	\ee
	lead to theories with equivalent thermodynamic variables. Thus starting from a  theory with general 4-derivative interactions, we can first perform the right field redefinitions to bring the Lagrangian into the form of Einstein-Maxwell-Weyl theory with coefficients depending on the couplings of the original theory. We then apply the improved RS method to calculate the on-shell Euclidean action. Finally, we recast the results in terms of the coefficients in the original theory. By this way,  we obtain thermodynamic variables for theories with general 4-derivative couplings without actually solving the corrected field equations. Surprisingly, this method also produces the right Casimir energy without invoking any surface terms associated with four derivative interactions.
	
	Upon choosing the parameters in the field redefinitions \eqref{field redefinition 1} to be
	\bea
	\lambda_1&=&-c_2-\frac{4c_3}{D-2}\,,\quad \lambda_2=\frac{4c_3}{(D-1)(D-2)}+\frac{c_2+2c_1}{D-2}\,,\quad
	\lambda_3=-\frac{1}{2}c_2-c_5-\frac{2(c_3+2c_6)}{D-2}\,,\cr
	\lambda_4&=&\frac{D-4}{2(D-2)^2}c_1+\frac{D-3}{2(D-2)^2}c_2
	+\frac{2D-5}{(D-2)^2(D-1)}c_3+\frac{2c_4+c_5}{D-2}+\frac{4c_6}{(D-2)(D-1)}\,,
	\eea
	the Einstein-Maxwell theory extended by general 4-derivative interactions \eqref{EMAction} reduces to the Einstein-Maxwell-Weyl theory \eqref{WeylAction} with parameters given by
	\begin{align}
		\widetilde{I}_{EMW} & =\frac{1}{16\pi}\int d^{D}x\sqrt{-g}(\mathcal{\widetilde{L}}_{EM}+\mathcal{\widetilde{L}}_{C^{2}})
		+\widetilde{I}_{{\rm GHY}}-\widetilde{I}_{{\rm ct}}\,,
		\nn\\
		\mathcal{\widetilde{L}}_{EM} & =\widetilde{\sigma}\big(R+\frac{(D-1)(D-2)}{\widetilde{\ell}_{0}^{2}}\big)
		-\frac{1}{4\widetilde{g}^{2}}F^{2}\,,\nn\\
		\mathcal{\widetilde{L}}_{C^{2}} & =\widetilde{\sigma}^{2}\widetilde{c}_{3}C^{\mu\nu\rho\sigma}C_{\mu\nu\rho\sigma}
		+\frac{\widetilde{\sigma}\widetilde{c}_{6}}{\widetilde{g}^{2}}C^{\mu\nu\rho\sigma}
		F_{\mu\nu}F_{\rho\sigma}+\frac{\widetilde{c}_{7}}{\widetilde{g}^{4}}(F^{2})^{2}
		+\frac{\widetilde{c}_{8}}{\widetilde{g}^{4}}
		F^{\mu}{}_{\nu}F^{\nu}{}_{\rho}F^{\rho}{}_{\sigma}F^{\sigma}{}_{\mu}\,.
		\label{Weyl 3}
	\end{align}
	where the GHY boundary term and surface counterterms  are
	\bea
	\widetilde{I}_{\mathrm{GHY}}&=&\frac{\widetilde{\sigma}}{8\pi }\int d^{D-1}x\sqrt{-h}K,\ \ \ \widetilde{I}_{\mathrm{ct}}=\frac{\widetilde{\sigma}}{16\pi }\int d^{D-1}x\sqrt{-h}\widetilde{\mathcal{L}}_{\mathrm{ct}}\,,
	\cr
	\widetilde{\mathcal{L}}_{\mathrm{ct}}&=&\frac{2(D-2) }{\widetilde{\ell}_{0}}
	+\frac{\widetilde{\ell}_{0}}{D-3}  \mathcal{R}+\frac{\widetilde{\ell}_{0}^3}{(D-5)(D-3)^2}  \big(\mathcal{R}_{ab}^2-\frac{D-1}{4(D-2)}\mathcal{R}^2\big)+\cdots\,.
	\eea
	Various coupling constants are renormalized by the field redefinitions and they take the form
	\bea
	\widetilde{ \sigma}&=&\sigma\Big(
	1-\frac{1}{2} (D-2) \lambda _0-\frac{\sigma  }{\ell_{0}^2}\big( (D-1) Dc_1+(D-1)c_2 +2 c_3\big)
	\Big),\cr
	\widetilde{\ell}_{0}&=&\ell_{0}\Big(1+\frac{\lambda _0}{2}-\frac{\sigma  }{2 \ell_{0}^2}\big( (D-1) Dc_1+ (D-1)c_2+2 c_3\big)\Big),\cr
	\frac{1}{\widetilde{g}^2}&=&\frac{1}{g^2}\Big[1+\frac{(D-4) \sigma }{(D-2) \ell_{0}^2}\big( (D-1) Dc_1+(D-1)c_2 +2 c_3\big)\cr
	&&+\frac{4 \sigma }{\ell_{0}^2}\big( (D-1) Dc_4+(D-1)c_5 +2 c_6\big)-\frac{1}{2} (D-4) \lambda _0
	\Big],\cr
	\widetilde{c}_3&=&c_3\,,\ \ \  \widetilde{c}_6=c_6\,,\ \ \ \widetilde{c}_8=c_8+\frac{1}{4} \left(c_2+2 c_5\right)+\frac{c_3+2 c_6}{D-2}\,, \cr
	\widetilde{c}_7&=&c_7+\frac{ (D-4)^2}{16 (D-2)^2}c_1-\frac{ (3 D-8)}{16 (D-2)^2}c_2-\frac{ (7 D-16)}{8 (D-2)^2 (D-1)}c_3\cr
	&&+\frac{ (D-4)}{4 (D-2)}c_4-\frac{c_5}{4 (D-2)}-\frac{3 c_6}{2 (D-2) (D-1)}\,.\label{expand tilde}
	\eea
	We now introduce the shorthand notation $\widetilde{\a}_{2,i}=(\widetilde{\sigma},\widetilde{\ell}_0,\widetilde{g})$, $\widetilde{\a}_{4,i}=(\widetilde{c}_3,\widetilde{c}_6,\widetilde{c}_7,\widetilde{c}_8)$ for simplicity. For Einstein-Maxwell-Weyl theory, treating $\widetilde{\alpha}_{4,i}$ as small parameters, we can employ the improved RS method to compute corrections to the leading order on-shell action. The results amount to replace all the coefficients in \eqref{Weyl 2 RS} with ones with tilde. We also put tilde on top of the temperature and electric potential to emphasize that the on-shell action is obtained for the Einstein-Maxwell-Weyl theory resulting from field redefinitions of the general theory. The result is given by
	\be
	\widetilde{T}\widetilde{I}_{ EMW}^{\rm RS}(\widetilde{T},\widetilde{\Phi}_e,\widetilde{\a}_{2,i},\widetilde{\a}_{4,i})
	=\widetilde{G}_{EMW}^{\rm RS}+\text{mod}[D,2]\times\widetilde{a}_{EMW}\,,
	\ee
	where
	\begin{align}
		&\widetilde{T}=\frac{(D-3)}{4\pi\widetilde{r}_{0}}\left(k
		-\frac{\widetilde{q}^{2}}{\widetilde{r}_{0}^{2(D-3)}}
		+\frac{(D-1)\widetilde{r}_{0}^{2}}{(D-3)\widetilde{\ell}_{0}^{2}}\right),
		\qquad\widetilde{\Phi}_{e}=\sqrt{\frac{2(D-2)\widetilde{\sigma}}{D-3}}
		\frac{\widetilde{g}\widetilde{q}}{\widetilde{r}_{0}^{D-3}}\,,
		\\
		&\widetilde{G}^{\rm RS}_{ EMW}  (\widetilde{T},\widetilde{\Phi}_{e},\widetilde{\a}_{2,i},\widetilde{\a}_{4,i})
		=-\frac{\widetilde{\sigma}\omega_{D-2,k}}{16\pi\widetilde{\ell}_{0}^{2}
			\widetilde{r}_{0}^{D-3}}\left(\widetilde{r}_{0}^{2(D-3)}
		(\widetilde{r}_{0}^{2}-k\widetilde{\ell}_{0}^{2})+\widetilde{q}^{2}
		\widetilde{\ell}_{0}^{2}\right)+\Delta\widetilde{G}^{\rm RS}_{ EMW}  \,,\nonumber \\
		&\Delta\widetilde{G}^{\rm RS}_{ EMW}    =\frac{(D-3)(D-2)\widetilde{\sigma}^{2}\omega_{D-2,k}}{16\pi(D-1)(3D-7)
			\widetilde{\ell}_{0}^{4}\widetilde{r}_{0}^{3D-7}}\Big[-(D-2)(D-1)(3D-7)
		\widetilde{r}_{0}^{4(D-3)}(k\widetilde{\ell}_{0}^{2}+\widetilde{r}_{0}^{2})^{2}
		\widetilde{c}_{3}
		\nonumber \\
		& +2(D-3)(D-1)(3D-7)\widetilde{q}^{2}\widetilde{\ell}_{0}^{2}
		\widetilde{r}_{0}^{2(D-3)}(k\widetilde{\ell}_{0}^{2}
		+\widetilde{r}_{0}^{2})(\widetilde{c}_{3}+\widetilde{c}_{6})
		-\big((D-3)(7D-16)\widetilde{c}_{3}
		\nonumber \\
		& +2(D-3)(5D-11)\widetilde{c}_{6}+8(D-2)(D-1)(2\widetilde{c}_{7}
		+\widetilde{c}_{8})\big)(D-3)\widetilde{q}^{4}\widetilde{\ell}_{0}^{4}\Big]\,,
	\end{align}
	and the Casimir energy $\widetilde{a}_{EMW}$ appearing in odd spacetime dimensions takes the form
	\bea
	\widetilde{a}_{EMW}=\frac{2(-k)^{\frac{1}{2}(D-1)}[(D-2)!!]^{2}}{(D-1)!}
	\frac{\widetilde{a}^{*}}{\widetilde{\ell}_{0}}\,, \qquad \widetilde{a}^{*}=\frac{\widetilde{\sigma}\omega_{D-2,k}}{16\pi}
	\widetilde{\ell}_0^{D-2}\,.
	\label{casimir2}
	\eea
	Notice that after expressing the parameters with tilde in terms of the parameters of the original theory according to \eqref{expand tilde}, the temperature and electric potential ($\widetilde{T},\,\widetilde{\Phi}_e$) will depend on coefficients of 4-derivative couplings $c_{i}$ explicitly. Hence we need to redefine the integration constants $\widetilde{r}_0$ and $\widetilde{q}$ to remove the explicit $c_{i}$s dependence
	\be
	\widetilde{r}_0=r_0+\delta r_0\,,\quad \widetilde{q}=q+\delta q\,,
	\label{rq2}
	\ee
	where
	\bea
	\delta r_{0} & =&\frac{r_{0}}{(D-2)\ell_{0}^{2}((D-3)\ell_{0}^{2}(q^{2}-kr_{0}^{2(D-3)})
		+(D-1)r_{0}^{2(D-2)})}
	\nonumber \\
	&& \times  \Bigg(-(D-2)(D-1)r_{0}^{2(D-2)}\big(-\lambda_{0}
	\ell_{0}^{2}+(D-1)\sigma(Dc_{1}+c_{2})+2\sigma c_{3}\big)
	\nonumber \\
	&& +\Big((D-2)\lambda_{0}\ell_{0}^{2}+2(D-3)(D-1)\sigma(Dc_{1}+c_{2})+8(D-2)\sigma c_{6}
	\nonumber \\
	&&+4(D-2)(D-1)\sigma(Dc_{4}+c_{5})\Big)(D-3)q^{2}\ell_{0}^{2}\Bigg),
	\cr
	\delta q & =&\frac{1}{2(D-2)\ell_{0}^{2}\big((D-3)\ell_{0}^{2}(q^{2}-kr_{0}^{2(D-3)})
		+(D-1)r_{0}^{2(D-2)}\big)}
	\nonumber \\
	&& \times  \Bigg((D-3)q\ell_{0}^{2}\big((2D-5)q^{2}-kr_{0}^{2(D-3)}\big)
	\Big[(D-2)\lambda_{0}\ell_{0}^{2}+4(D-3)\sigma c_{3}
	\nonumber \\
	&&+2(D-3)(D-1)\sigma(Dc_{1}+c_{2})+4(D-2)(D-1)\sigma\left(Dc_{4}
	+c_{5}\right)+8(D-2)\sigma c_{6}\Big]
	\nonumber \\
	&&+(D-1)qr_{0}^{2(D-2)}\Big[(D-2)(2D-5)\lambda_{0}\ell_{0}^{2}-4(D-3)^{2}\sigma c_{3}+8(D-2)\sigma c_{6}
	\nonumber \\
	&& -2(D-1)(D-3)^{2}\sigma(Dc_{1}+c_{2})+4(D-2)(D-1)\sigma(Dc_{4}+c_{5})\Big]\Bigg).
	\eea
	Taking into account \eqref{rq2}, we finally obtain the on-shell action for static charged AdS black holes in Einstein-Maxwell theory extended by general 4-derivative interactions. Denoting  this model by ``EMG", we have
	\bea
	&&TI_{EMG}=G_{EMG}+\text{mod}[D,2]\times a_{EMG}\,,
	\label{EMG}
	\eea
	where
	\begingroup\allowdisplaybreaks
	\begin{align}
		&T=\frac{D-3}{4\pi r_{0}}\Big(k-\frac{q^{2}}{r_{0}^{2(D-3)}}+\frac{(D-1)r_{0}^{2}}{(D-3)
			\ell_{0}^{2}}\Big)\,, \qquad \Phi_{e}=-\sqrt{\frac{2(D-2)\sigma}{D-3}}\frac{gq}{r_{0}^{D-3}}\,,
		\nn\\
		&G_{EMG}  (T,\Phi_{e},c_{i})=-\frac{\sigma\omega_{D-2,k}}{16\pi\ell_{0}^{2}r_{0}^{D-3}}
		\left(r_{0}^{2(D-3)}(r_{0}^{2}-k\ell_{0}^{2})+q^{2}\ell_{0}^{2}\right)+\Delta G_{EMG}\,,\cr
		&\Delta G_{EMG}  =\frac{\sigma r_{0}^{7-3D}\omega_{D-2,k}}{32\pi(3D-7)\ell_{0}^{4}}\Bigg(4\left(c_{3}
		+c_{6}\right)(D-3)^{2}\left(3D^{2}-13D+14\right)\sigma kq^{2}\ell_{0}^{4}r_{0}^{2(D-3)}\nonumber \\
		& -2c_{3}(D-3)(D-2)^{2}(3D-7)\sigma k^{2}\ell_{0}^{4}r_{0}^{4(D-3)}-(3D-7)k\ell_{0}^{2}r_{0}^{2(2D-5)}
		\Big[(D-2)\lambda_{0}\ell_{0}^{2}\nonumber \\
		& +2(D-1)\sigma\left(Dc_{1}+c_{2}\right)+4c_{3}\left(D^{3}-7D^{2}+16D-11\right)
		\sigma\Big]+\Big[-(D-2)\lambda_{0}\ell_{0}^{2}\nonumber \\
		& +2(D-1)^{2}\sigma\left(Dc_{1}+c_{2}\right)-2c_{3}\left(D^{3}
		-7D^{2}+14D-10\right)\sigma\Big](3D-7)r_{0}^{4(D-2)}\nonumber \\
		& -2(D-3)^{2}\sigma q^{4}\ell_{0}^{4}\Big[(D-4)^{2}c_{1}+(2D^{2}-11D+16)c_{2}+(7D^{2}-36D+48)c_{3}
		\nonumber \\
		& +4(D-2)(D-4)c_{4}+4(D-3)(D-2)c_{5}+2(D-2)(5D-13)c_{6}\nonumber \\
		& +8(D-2)^{2}(2c_{7}+c_{8})\Big]-(3D-7)q^{2}\ell_{0}^{2}r_{0}^{2(D-2)}
		\Big[+(D-2)\lambda_{0}\ell_{0}^{2}\nonumber \\
		& +2(D-1)(2D-7)\sigma(Dc_{1}+c_{2})-4(D^{3}-8D^{2}+19D-11)c_{3}\sigma\nonumber \\
		& +8(D-2)(D-1)\sigma(Dc_{4}+c_{5})-4c_{6}(D-5)(D-2)(D-1)\sigma\Big]\Bigg).
	\end{align}\endgroup
	The effective AdS radius for the model with general 4-derivative couplings is related to the bare AdS radius via
	\be
	\ell_{0}=\ell_{\rm eff}+\frac{(D-4)\sigma}{2(D-2)\ell_{\rm eff}}\big(D(D-1)c_1+(D-1)c_2+2c_3
	\big) \,,
	\ee
	using which, the Casimir energy term in \eqref{EMG} can be rewritten as
	\be
	a_{EMG}=\frac{2(-k)^{\frac{1}{2}(D-1)}[(D-2)!!]^2}{(D-1)!}
	\frac{\widetilde{a}^*}{\widetilde{\ell}_0}=\frac{2(-k)^{\frac{1}{2}(D-1)}[(D-2)!!]^2}{(D-1)!}\frac{a}{\ell_{\rm eff}}\,,
	\ee
	in which
	\be
	a=\frac{\sigma\omega_{D-2,k}}{16\pi }\Big(\ell_{\rm eff}^{D-2}-2\sigma\big(D(D-1)c_1+(D-1)c_2+2c_3\big)\ell_{\rm eff}^{D-4}
	\Big)\,.
	\ee
	This is precisely the $a$ central charge for the model with general 4-derivative couplings.
	
	Using results given in section 3, we have also checked that the results obtained by improved RS method agree with those obtained by the ordinary method of solving the linear-order perturbative solution explicitly. The perturbative solutions are given in appendix \ref{Perturbed solution}.
	
\section{Conclusions}
\label{sec:con}

The original proposal of RS for AdS black holes (see \cite[footnote 6]{Reall:2019sah}) could give the correct answer as long as the induced metric on the conformal boundary is preserved. However, in practice one needs to solve for the modified AdS radius and construct the explicit form of the surface terms associated with the bulk higher derivative interactions. If the former is an easy task, the latter is highly nontrivial. This is because that the surface terms include both the GHY terms and holographic counterterms. Although these terms are known for Lovelock theories including the GB terms in particular, they are not known for general curvature invariants based on Riemann tensor the its covariant derivatives. In fact the explicit forms of these surface terms can be rather non-trivial to work out for a generic higher-derivative bulk term.  Recently an enhanced RS procedure was proposed \cite{Xiao:2023two}, by using the background subtraction. This method does not require the surface terms, however it is known to be ambiguous for stationary rotating black holes and may lead to incorrect result. The method of subtracting the vacuum background can also be problematic for topological AdS black holes with hyperbolic horizon since the vacuum itself is already thermal.

In this paper, we resolved these issues by proposing an improvement of the RS method for AdS black holes that does not require the construction of any higher derivative surface terms. This greatly simplifies the RS method
applied to AdS black holes as  one only needs to substitute the leading-order solution into the bulk action. Thus our approach drastically simplifies computing quantum corrections to thermodynamic valuables of general AdS black holes and possess numerous applications in AdS quantum gravity.  In particular, it facilities the precision test of AdS/CFT correspondence \cite{Bobev:2022bjm, Cassani:2022lrk, Ma:2024ynp}.
To achieve this, we first observed that the RS method gave the correct answer for Einstein-Maxwell-Weyl theory which does not modify the AdS radius and requires no higher derivative surface terms. Then starting from a theory with general 4-derivative interactions, we performed proper field redefinitions to bring the Lagrangian into the form of Einstein-Maxwell-Weyl theory for which the RS method is applied to compute  the on-shell action of the static charged AdS black hole solutions. Finally, the results are reexpressed in terms of the coefficients of the original theory. By this way,  we obtain thermodynamic variables for AdS black holes in theories with general 4-derivative couplings without actually solving the corrected field equations.
	We verified our results by constructing explicitly the leading-order perturbative solutions. Although we focused on the spherically-symmetric and static charged AdS black holes, the appropriate thermodynamically-equivalent field redefinitions we obtained for the general 4-derivative extensions are valid for all the black hole solutions in the Einstein-Maxwell theories.
	
	Surprisingly, our method also produces the right Casimir energy without invoking any surface terms associated with four derivative interactions. As we can see in appendix C, our improved RS method provides a quick way to derive how the higher-derivative corrections affect the mass-charge and entropy-charge relations. Requiring $\Delta S_{\rm ext}\ge 0$ from higher-derivative terms imposes inequality constraints on their coupling constants, namely
	\bea
	X&\equiv&c_2+2 c_3+2 c_5+4 c_6+16 c_7+8 c_8\le 0\,,\cr
	Y&\equiv &2 c_1+c_2+2 c_3+4 c_4+2 c_5+4 c_6+8 c_7+4 c_8\le0\,,\cr
	Z&\equiv& c_2+2 c_3+4 c_4+3 c_5+6 c_6+16 c_7+8 c_8\le
	2\sqrt{ XY}\,.\label{positivecond}
	\eea
	
	As for future directions, we would like to apply our method to more general AdS black holes. In the context of precision test of AdS/CFT correspondence, it should be very interesting to perform a complete analysis of 4-derivative corrections to the Gutowski-Reall solution \cite{Gutowski:2004yv} in 5D $N=2$ supergravities based on the recent construction of all the gauged curvature squared invariants \cite{Ozkan:2013uk,Ozkan:2013nwa,Butter:2014xxa,Gold:2023dfe,Gold:2023ymc,Gold:2023ykx},
	and study whether the higher-derivative terms could contribute negatively to the black hole entropy, as in the case of \cite{Ma:2022gtm}. Another direction is to extend this leading-order perturbation to even higher-order corrections, generalizing the asymptotically-flat results of \cite{Ma:2023qqj} to AdS black holes. We expect that this can be principally done since one can always perform field redefinitions to rewrite the higher-derivative corrections to the Weyl structures.
	
%

	\section*{Acknowlegment}
	
	We would like to thank D. Cassani, A. Ruiprez, E. Turetta for useful discussions. We also thank the anonymous referee whose comments help us improve the paper significantly.
	L.M.~would like to thank Peng Huanwu Center for Fundamental Theory, Hefei for hospitality during the early stage of this work. The work of H.L. is supported in part by the National Natural Science Foundation of China (NSFC) grants No. 11935009 and No. 12375052. This work of Y.P. is supported by the National Key Research and Development Program under grant No. 2022YFE0134300 and National Natural Science Foundation of China (NSFC) No. 12175164.

	\appendix
	
		\section{Naive application of RS method to AdS black hole in EGB gravity}\label{AppendixA}
		Up to first order in $\alpha$, we evaluate  the on-shell Euclidean action \eqref{Lct} of AdS black hole in EGB theory directly  by using the uncorrected solution. The result is given by
		\be
		T' I'_{E}= F'+\text{mod}[D,2]\times (a_{1}+a_{2})\,,\qquad T'=\frac{D-3}{4 \pi  r_{0}'}\Big(k+\frac{(D-1) r_{0}'^2}{(D-3) \ell_{0}^2}\Big)\ ,
		\label{I0S}
		\ee
		where $a_1$ is defined in \eqref{a1}, $F'$ and $a_2$ are given below
		\bea
		F' &=&\frac{r_{0}'^{D-3}\left(k\ell_{0}^{2}-r_{0}'^{2}\right) \omega_{D-2,k}}{16\pi\ell_{0}^{2}G_{D}}
		\nn\\
		&&-\frac{\alpha(D-3)(D-2)r_{0}'^{D-5}\left(2(D-2)k^{2}\ell_{0}^{4} +(5D-8)kr_{0}'^{2}\ell_{0}^{2}+(D-4)r_{0}'^{4}\right) \omega_{D-2,k}}{32\pi\ell_{0}^{4}G_{D}}\,,\nn \\
		a_{2}&=&-\alpha\frac{[(D-2)\text{!!}]^{2}(-k)^{\frac{D-1}{2}}}{(D-1)! \ell_{0}}\frac{(D-4)(D-3)\ell_{0}^{D-4}\omega_{D-2,k}}{16\pi G_{D}}\,,
		\eea
		To compare \eqref{IBR} with \eqref{I0S}, we set the parameters $\tr_0=r_0'$ so that these two Euclidean actions are defined at the same temperature. We then find that
		\be
		T(I_{E}-I_{E}')=\alpha\frac{(D-4)(D-3)(D-2)r_{0}'^{D-3}\left(k\ell_{0}^{2} +r_{0}'^{2}\right)\omega_{D-2,k}}{32\pi\ell_{0}^{4}G_{D}}-{\rm mod}[D,2]\times a_{2}\,,\label{dis}
		\ee
		which clearly states that the Euclidean action derived  directly  from the uncorrected solution is different from the one obtained by using the solution with 4-derivative corrections.

	\section{Perturbed solutions}\label{Perturbed solution}
	
	The 4-derivative terms modify the field equations and thus the solutions. Up to first order in $c_i$, the corrections \eqref{perturbed solution3} to the solution of the 2-derivative theory, with parameters $(\mu,q)$ fixed,  are given by
	\begingroup\allowdisplaybreaks
	\begin{align}
		\Delta h & =\frac{2(D-3)\sigma}{(D-2)r^{2(D-2)}}\Bigg(2(D-4)(D-2)\mu^{2}c_{3}+
		kq^{2}\Big(2(D-4)c_{1}-\left(D^{2}-6D+10\right)c_{2}\nonumber \\
		& -2\left(2D^{2}-11D+16\right)c_{3}+4(D-2)c_{4}-
		(D-4)(D-2)c_{5}-2(D-3)(D-2)c_{6}\Big)\Bigg)\nonumber \\
		& -\frac{4(D-3)\sigma\mu q^{2}}{(D-2)r^{3D-7}}\Big((D-4)(D-1)c_{1}-c_{2}-Dc_{3}+2(D-2)(D-1)c_{4}
		+(D-2)c_{5}\nonumber \\
		& -(D-3)(D-2)c_{6}\Big)+\frac{(D-3)\sigma q^{4}}{(D-2)(3D-7)r^{2(2D-5)}}\Big((D-4)\left(11D^{2}-45D+44\right)
		c_{1}\nonumber \\
		& +\left(4D^{3}-33D^{2}+83D-64\right)c_{2}+2(D-2)\left(D^{2}-D-4\right)c_{5}
		-4(D-2)(D-3)^{2}c_{6}\nonumber \\
		& +2\left(4D^{3}-34D^{2}+87D-68\right)c_{3}-16(D-2)^{2}(D-3)c_{7}
		-8(D-2)^{2}(D-3)c_{8}\nonumber \\
		& +4(D-2)\left(5D^{2}-19D+16\right)c_{4}\Big)-\frac{2\sigma q^{2}}{\ell_{0}^{2}r^{2(D-3)}}\big((1-D)Dc_{1}+
		\left(D^{2}-6D+7\right)c_{2}\nonumber \\
		& +2(D-1)Dc_{4}+2\left(2D^{2}-9D+8\right)c_{3}+\left(D^{2}-3D+4\right)c_{5}
		+2\left(D^{2}-4D+5\right)c_{6}\big)\nonumber \\
		& +\frac{(D-4)\sigma r^{2}}{(D-2)\ell_{0}^{4}}\big((D-1)Dc_{1}+(D-1)c_{2}+2c_{3}\big),\nonumber \\
		\Delta f & =\Delta h-\frac{4(D-3)\sigma q^{2}f_{0}}{(D-2)r^{2(D-2)}}\Big((D-4)(2D-3)c_{1}+(D-2)(D-1)(c_{5}+c_{6})\nonumber \\
		& +\left(D^{2}-5D+5\right)c_{2}+\left(2D^{2}-9D+8\right)c_{3}
		+2(D-2)(2D-3)c_{4}\Big),\nonumber \\
		\Delta\psi & =\frac{2\sqrt{2}(D-3)^{3/2}\sigma^{3/2}gq^{3}}{\sqrt{D-2}(3D-7)r^{3D-7}}
		\Big(-(7D-19)(D-2)c_{6}-8(D-2)^{2}(2c_{7}+c_{8})\nonumber \\
		& +\left(D^{2}-5D+5\right)c_{2}+\left(2D^{2}-9D+8\right)c_{3}
		+(D-2)[2(D+1)c_{4}-(D-5)c_{5}]\nonumber \\
		& +(D-4)(2D-3)c_{1}\Big)-
		\frac{4\sqrt{2(D-2)}\sigma^{3/2}gq}{\sqrt{D-3}\ell_{0}^{2}r^{D-3}}
		\big((D-1)Dc_{4}+(D-1)c_{5}+2c_{6}\big)\nonumber \\
		& +\frac{4\sqrt{2}(D-3)^{3/2}\sqrt{D-2}\sigma^{3/2}\mu gq}{r^{2(D-2)}}c_{6}\,.
	\end{align}
	\endgroup
	Under the perturbation with fixed $(\mu,q)$, the radius of black hole horizon is shifted to be $r_h=r_{0}+\Delta r$ where
	\begingroup\allowdisplaybreaks
	\begin{align}
		\Delta r & =\frac{\sigma}{(D-2)(3D-7)\ell_{0}^{2}r_{0}^{2D-5}\big((D-3)k\ell_{0}^{2}r_{0}^{2(D-3)}
			+(D-1)r_{0}^{2(D-2)}-(D-3)q^{2}\ell_{0}^{2}\big)}\nonumber \\
		& \times\Big(2(3D-7)q^{2}\ell_{0}^{2}r_{0}^{2(D-2)}\big(-(D-1)(5D-12)c_{1}+
		\left(D^{2}-7D+11\right)(D-1)c_{2}\nonumber \\
		& +\left(3D^{3}-18D^{2}+29D-8\right)c_{3}+(D-2)(D-1)^{2}c_{5}+
		(D-2)(D-1)^{2}c_{6}\nonumber \\
		& +2(D-2)(2D-3)(D-1)c_{4}\big)-(D-3)^{2}(D-2)q^{4}\ell_{0}^{4}
		\big(-8(D-2)(2c_{7}+c_{8})\nonumber \\
		& +5(D-4)c_{1}+(4D-13)c_{2}+(11D-32)c_{3}+4(2D-3)c_{4}
		+2(D-1)(c_{5}+c_{6})\big)\nonumber \\
		& -c_{3}(D-4)(D-3)(D-2)(3D-7)(r_{0}^{4(D-2)}+2k\ell_{0}^{2}r_{0}^{2(2D-5)}
		+k^{2}\ell_{0}^{4}r_{0}^{4(D-3)})\nonumber \\
		& +2(D-3)^{2}(3D-7)kq^{2}\ell_{0}^{4}r_{0}^{2(D-3)}
		\big((D-4)c_{1}+(D-3)c_{2}+(D-2)(2c_{4}+c_{5}+c_{6})\nonumber \\
		& +(3D-8)c_{3}\big)\Big)+\frac{(D-4)\sigma r_{0}^{2D-3}\big((D-1)(Dc_{1}+c_{2})+2c_{3}\big)}{(D-2)\ell_{0}^{2}
			\big((D-3)k\ell_{0}^{2}r_{0}^{2(D-3)}+(D-1)r_{0}^{2(D-2)}
			-(D-3)q^{2}\ell_{0}^{2}\big)}\,.
	\end{align}\endgroup
	
	\section{Corrections to thermodynamic variables}
	\label{Corrected Thermodynamics}
	
	For the general 4-derivative extensions of the Einstein-Maxwell theory, the mass and electric charges are given in \eqref{mq}. The rest of thermodynamic variables are listed below.
	Up to first order in $c_i$, the  entropy is $S  =S_{0}+\Delta S$ with
	\begingroup\allowdisplaybreaks
	\begin{align}
		\Delta S & =\frac{\omega_{D-2,k}\sigma^{2}}{4(3D-7)\ell_{0}^{2}r_{0}^{D-2}
			\left((D-3)k\ell_{0}^{2}r_{0}^{2(D-3)}+(D-1)r_{0}^{2(D-2)}
			-(D-3)q^{2}\ell_{0}^{2}\right)}\nonumber \\
		& \times\Bigg((D-3)^{2}q^{4}\ell_{0}^{4}\Big((D-4)^{2}c_{1}
		+\left(7D^{2}-36D+48\right)c_{3}+8(D-2)^{2}\left(2c_{7}+c_{8}\right)\nonumber \\
		& +\left(2D^{2}-11D+16\right)c_{2}+4(D-2)((D-4)c_{4}+(D-3)c_{5})
		+2(D-2)(5D-13)c_{6}\Big)\nonumber \\
		& +(D-3)(D-2)(3D-7)k\ell_{0}^{4}\left((D-2)kc_{3}r_{0}^{4(D-3)}
		-2(D-3)\left(c_{3}+c_{6}\right)q^{2}r_{0}^{2(D-3)}\right)\nonumber \\
		& -2(3D-7)q^{2}\ell_{0}^{2}r_{0}^{2(D-2)}\Big(-2(D-2)(D-1)
		\left(Dc_{4}+c_{5}\right)+(D-5)(D-2)(D-1)c_{6}\nonumber \\
		& +(D-1)\left(Dc_{1}+c_{2}\right)+(D-4)(D^{2}-6D+7)c_{3}\Big)
		-(3D-7)\Big(2(D-1)^{2}(Dc_{1}+c_{2})\nonumber \\
		& -(D-4)(D^{2}+D-4)c_{3}\Big)r_{0}^{4(D-2)}
		-2(D-3)(3D-7)k\ell_{0}^{2}r_{0}^{2(2D-5)}
		\Big((D-1)\left(Dc_{1}+c_{2}\right)\nonumber \\
		& -\left(D^{2}-2D-2\right)c_{3}\Big)\Bigg)
		-\frac{(D-4)\sigma^{2}r_{0}^{3(D-2)}\omega_{D-2,k}
			\big((D-1)(c_{1}D+c_{2})+2c_{3}\big)}{4\ell_{0}^{2}
			\big((D-3)k\ell_{0}^{2}r_{0}^{2(D-3)}+(D-1)r_{0}^{2(D-2)}
			-(D-3)q^{2}\ell_{0}^{2}\big)}\,.
	\end{align}\endgroup
	The temperature is modified to be $T =T_{0}+\Delta T$ with
	\begingroup\allowdisplaybreaks
	\begin{align}
		\Delta T & =\frac{\sigma}{4\pi(D-2)(3D-7)\ell_{0}^{4}r_{0}^{4D-9}
			\big((D-3)k\ell_{0}^{2}r_{0}^{2(D-3)}+(D-1)r_{0}^{2(D-2)}
			-(D-3)q^{2}\ell_{0}^{2}\big)}\nonumber \\
		& \times\Bigg(-(D-3)^{3}(D-2)q^{6}\ell_{0}^{6}\Big((D-4)^{2}c_{1}+(2D^{2}-11D+16)c_{2}
		+(7D^{2}-36D+48)c_{3}\nonumber \\
		& +4(D-2)\left((D-4)c_{4}+(D-3)c_{5}\right)+2(D-2)(5D-13)c_{6}
		+8(D-2)^{2}(2c_{7}+c_{8})\Big)\nonumber \\
		& +(D-3)^{3}kq^{4}\ell_{0}^{6}r_{0}^{2(D-3)}\Big(8(3D^{2}-15D+19)(D-2)c_{6}
		+8(3D-8)(D-2)^{2}(2c_{7}+c_{8})\nonumber \\
		& +(15D^{3}-114D^{2}+300D-272)c_{3}+4(D-2)(3D-8)\left((D-4)c_{4}
		+(D-3)c_{5}\right)\nonumber \\
		& +(D-4)^{2}(3D-8)c_{1}+(3D-8)(2D^{2}-11D+16)c_{2}\Big)
		+k^{2}q^{2}\ell_{0}^{6}r_{0}^{4(D-3)}(D-3)^{2}(D-2)^{2}\nonumber \\
		& \times(3D-7)\left((D-4)c_{3}-2(D-3)c_{6}\right)
		-(D-4)(D-3)^{2}(D-2)^{2}(3D-7)k^{3}c_{3}\ell_{0}^{6}r_{0}^{6(D-3)}\nonumber \\
		& +(D-3)(D-2)\times\Big(4(D-1)(D-2)\left[2(3D^{2}-14D+18)c_{4}
		+(3D^{2}-15D+20)c_{5}\right]\nonumber \\
		& -(D-1)\left(3D^{3}+7D^{2}-92D+144\right)c_{1}+(D-1)
		\left(6D^{3}-57D^{2}+173D-172\right)c_{2}\nonumber \\
		& +\left(15D^{4}-106D^{3}+201D^{2}+42D-296\right)c_{3}
		+4(D-1)(D-2)\left(6D^{2}-31D+41\right)c_{6}\nonumber \\
		& +24(D-3)(D-1)(D-2)^{2}(2c_{7}+c_{8})\Big)q^{4}\ell_{0}^{4}r_{0}^{2(D-2)}
		+(D-2)(3D-7)q^{2}\ell_{0}^{2}r_{0}^{4(D-2)}\nonumber \\
		& \times\Big(+2(D-2)(D-1)^{2}\left[-(Dc_{1}+c_{2})+2(Dc_{4}
		+c_{5})\right]+c_{3}(D-4)(D^{3}-15D+26)\nonumber \\
		& -2c_{6}(D-5)(D-2)(D-1)^{2}\Big)-(D-4)(D-3)(D-2)(3D-7)c_{3}r_{0}^{2(3D-8)}
		\Big((D-1)Dr_{0}^{4}\nonumber \\
		& +(3D^{2}\!-7D\!+6)k\ell_{0}^{2}r_{0}^{2}+(3D^{2}\!-\!11D\!+12)k^{2}\ell_{0}^{4}\Big)
		\!+\!2(D-3)(D-2)(3D-7)kq^{2}\ell_{0}^{4}r_{0}^{2(2D-5)}\nonumber \\
		& \times\Big(2(D-4)(D-1)(Dc_{4}+c_{5})-(D-4)(D-1)(Dc_{1}+c_{2})
		-2(D-1)(D^{2}-6D+10)c_{6}\nonumber \\
		& +(D-4)(D^{2}-D-8)c_{3}\Big)\Bigg)+(D-4)\sigma r_{0}\big((D-1)\left(Dc_{1}+c_{2}\right)+2c_{3}\big)\nonumber \\
		& \times\frac{(D-3)\ell_{0}^{2}\left(Dkr_{0}^{2(D-3)}-3(D-2)q^{2}\right)
			+(D-2)(D-1)r_{0}^{2(D-2)}}{4\pi(D-2)\ell_{0}^{4}
			\big((D-3)k\ell_{0}^{2}r_{0}^{2(D-3)}+(D-1)r_{0}^{2(D-2)}
			-(D-3)q^{2}\ell_{0}^{2}\big)}\,.
	\end{align}\endgroup
	Finally, the corrected  electric potential is $\Phi_{e}=\Phi_{e,0}+\Delta\Phi_{e}$ with
	\begingroup\allowdisplaybreaks
	\begin{align}
		\Delta\Phi_{e} & =\frac{-\sqrt{2}\sigma^{3/2}gq}{\sqrt{(D-3)(D-2)}(3D-7)
			\ell_{0}^{2}r_{0}^{3D-7}\left((D-3)(k\ell_{0}^{2}r_{0}^{2(D-3)}
			-q^{2}\ell_{0}^{2})+(D-1)r_{0}^{2(D-2)}\right)}\nonumber \\
		& \times\Bigg(2(D-3)(D-2)(3D-7)\cr
		&\times\left((D-4)(D-3)c_{3}-2(D-1)
		\left(Dc_{4}+c_{5}-(D-4)c_{6}\right)\right)k\ell_{0}^{2}r_{0}^{2(2D-5)}\nonumber \\
		& +(D-2)(3D-7)\left((D-4)(D-3)^{2}c_{3}+2(D-1)^{2}
		\left(-2Dc_{4}-2c_{5}+(D-5)c_{6}\right)\right)r_{0}^{4(D-2)}\nonumber \\
		& -2(D-3)q^{2}\ell_{0}^{2}r_{0}^{2(D-2)}\times
		\Big(+4(D-3)(D-1)(D-2)\left((D-4)c_{4}+(D-3)c_{5}\right)\nonumber \\
		& +2(D-3)(D-1)(5D-13)(D-2)c_{6}+8(D-3)(D-1)(D-2)^{2}(2c_{7}+c_{8})\nonumber \\
		& +2(D-1)(D^{3}-10D^{2}+31D-31)c_{2}+(7D^{4}-58D^{3}+163D^{2}-168D+32)c_{3}\nonumber \\
		& -2(D-1)(D^{3}-D^{2}-13D+24)c_{1}\Big)
		+(D-3)^{3}q^{2}\ell_{0}^{4}(q^{2}-2kr_{0}^{2D-6})\Big((D-4)^{2}c_{1}\nonumber \\
		& +(2D^{2}-11D+16)c_{2}+(7D^{2}-36D+48)c_{3}+4(D-2)\left((D-4)c_{4}
		+(D-3)c_{5}\right)\nonumber \\
		& +2(D-2)(5D-13)c_{6}+8(D-2)^{2}(2c_{7}+c_{8})\Big)
		+\left((D-4)c_{3}+2(D-3)c_{6}\right)k^{2}\ell_{0}^{4}r_{0}^{4(D-3)}\nonumber \\
		& \times(D-3)^{2}(D-2)(3D-7)\Bigg)\cr
		&
		-\frac{(D-4)\sqrt{2\sigma(D-3)}\sigma gqr_{0}^{D-1}\big((D-1)(Dc_{1}+c_{2})
			+2c_{3}\big)}{\sqrt{D-2}\ell_{0}^{2}\big((D-3)\ell_{0}^{2}(kr_{0}^{2(D-3)}
			-q^{2})+(D-1)r_{0}^{2(D-2)}\big)}\,.
	\end{align}
	\endgroup
	
	\section{(Mass-)entropy-charge relations in the extremal limit}
	
	The charged black holes contain two independent integration constants, the mass $M$ and charge $Q_e$. In the extremal limit, they become algebraically related and the general formula can be complicated. However, they can be both expressed parametrically in terms of the horizon the horizon radius $r_0$. To see how the higher-derivative correction affects the mass-charge and entropy-charge relations, we fix the charge. In the extremal limit, we have
	\be
	Q_{e}=\frac{\sqrt{(D-2) \sigma } \Omega _{D-2,k}}{8 \sqrt{2} \pi  g \ell _0}r_0^{D-3} \sqrt{(D-3) k \ell _0^2+(D-1) r_0^2}\,.\label{echargeext}
	\ee
	The mass and entropy are then affected by the higher-derivative terms, given by
	\bea
	&&M_{\mathrm{ext}}=\frac{(D-2) \sigma  r_0^{D-3} \Omega _{D-2,k}}{8 \pi  (D-3) \ell _0^2}\left(
	(D-3) k \ell _0^2+(D-2) r_0^2
	\right)-\frac{\sigma ^2 r_0^{D-5} \Omega _{D-2,k}}{16 \pi  (D-3) (3 D-7) \ell _0^4}\Big[\cr
	&&4 \left(c_5+c_6\right) (D-2) \left((D-3) k \ell _0^2+(D-1) r_0^2\right) \left((D-3)^3 k \ell _0^2+(D-2) (D-1)^2 r_0^2\right)\cr
	&&+c_1 (D-3)\big(
	4 (D-1) \left(2 D^3-12 D^2+27 D-24\right) k r_0^2 \ell _0^2\cr
	&&+4 (D-2) (D-1) \left(D^2-2 D+2\right) r_0^4 +(D-4)^2 (D-3)^2 k^2 \ell _0^4
	\big)\cr
	&& +8 \left(2 c_7+c_8\right) (D-3) (D-2)^2 \left((D-3) k \ell _0^2+(D-1) r_0^2\right)^2\cr
	&&+4 c_4 (D-2)\big(
	(D-3) (D-1) \left(5 D^2-21 D+24\right) k r_0^2 \ell _0^2+(D-4) (D-3)^3 k^2 \ell _0^4\cr
	&&+2 (D-2) (D-1)^2 (2 D-3) r_0^4
	\big)+c_2 (D-3)\big(
	(D-3)^2 \left(2 D^2-11 D+16\right) k^2 \ell _0^4\cr
	&&+4 (D-1) \left(D^3-7 D^2+18 D-17\right) k r_0^2 \ell _0^2+2 (D-2) (D-1) \left(D^2-3 D+4\right) r_0^4
	\big)\cr
	&&+2 c_3 (D-3)\big(
	\left(2 D^3-16 D^2+45 D-44\right) k \ell _0^2 \left((D-3) k \ell _0^2+2 (D-1) r_0^2\right)\cr
	&&+2 (D-2) (D-1) \left(D^2-4 D+6\right) r_0^4
	\big)
	\Big],\cr
	&&S_{\mathrm{ext}}=\frac{1}{4} \sigma  r_0^{D-2} \Omega _{D-2,k}-\frac{\sigma ^2 r_0^{D-4} \Omega _{D-2,k}}{8 \ell _0^2 \left((D-3)^2 k \ell _0^2+(D-2) (D-1) r_0^2\right)}\Big[\cr
	&&8 \left(2 c_7+c_8\right) (D-2)^2 \left((D-3) k \ell _0^2+(D-1) r_0^2\right)^2
	\cr
	&&+c_1 \left((D-4) (D-3) k \ell _0^2+2 (D-2) (D-1) r_0^2\right)^2\cr
	&&
	+4 \left(c_5+2 c_6\right) (D-2) \left((D-3) k \ell _0^2+(D-1) r_0^2\right) \left((D-3)^2 k \ell _0^2+(D-2) (D-1) r_0^2\right)\cr
	&&
	+4 c_4 (D-2) \big((D-4) (D-3)^2 k^2 \ell _0^4+(D-3) (D-1) (3 D-8) k r_0^2 \ell _0^2
	\cr
	&&+2 (D-2) (D-1)^2 r_0^4\big) +c_2 \big((D-3)^2 \left(2 D^2-11 D+16\right) k^2 \ell _0^4\cr
	&&+4 (D-3)^2 (D-2) (D-1) k r_0^2 \ell _0^2+2 (D-2)^2 (D-1)^2 r_0^4\big)\cr
	&&
	+c_3 \big(2 (D-3) \left(2 D^3-18 D^2+55 D-56\right) k^2 \ell _0^4+8 (D-3)^2 (D-2) (D-1) k r_0^2 \ell _0^2\cr
	&&+4 (D-2)^2 (D-1)^2 r_0^4\big)\Big].
	\eea
	We can solve for $r_0$ as a function of $Q$, and substitute the results into the mass and entropy expression, we arrive at $M(Q)$ and $S(Q)$ in the extremal limit.
	
	As an explicit example, we present the results for $D=4$ spherically symmetric extremal charged AdS black holes. In this case, we have $k=1$ and $\Omega_{2,1}=4\pi$, and therefore
	\be
	r_0^2= \frac{1}{6} \ell^2 _0 \left(\sqrt{{\cal Q}^2+1}-1\right),\qquad  {\cal Q} \equiv \sqrt{\fft{48 g^2}{\sigma \ell^2}}\, Q_e>0\,.
	\ee
	In terms of the dimensionless charge ${\cal Q}$, we have some rather simple relations despite complicated higher-derivative corrections:
	\bea
	M_{\mathrm{ext}}&=&\frac{\sqrt{\sqrt{\mathcal{Q}^2+1}-1} \left(\sqrt{\mathcal{Q}^2+1}+2\right) \sigma  \ell _0}{3 \sqrt{6}}+\frac{\sigma ^2}{5 \sqrt{6} \sqrt{\sqrt{\mathcal{Q}^2+1}-1} \ell _0}\Big[\cr
	&&(40 c_1+7 c_2-12 c_3-6 c_5-6 c_6-48 c_7-24 c_8)\cr
	&&-2 \left(5 c_1+2 c_2+3 c_3+30 c_4+9 c_5+9 c_6+12 c_7+6 c_8\right) \mathcal{Q}^2\cr
	&&-\sqrt{\mathcal{Q}^2+1}\left(40 c_1+13 c_2+12 c_3+6 c_5+6 c_6+48 c_7+24 c_8\right)
	\Big],\cr
	S_{\mathrm{ext}}&=&\frac{\pi  \left(\sqrt{\mathcal{Q}^2+1}-1\right) \sigma  \ell _0^2}{6} -\frac{\pi  \sigma ^2}{\sqrt{\mathcal{Q}^2+1}}\Big[
	2 \left(2 c_1+c_2+2 c_3+c_5+2 c_6+8 c_7+4 c_8\right)\cr
	&&+\left(2 c_1+c_2+2 c_3+4 c_4+2 c_5+4 c_6+8 c_7+4 c_8\right) \mathcal{Q}^2\cr
	&&-2 \sqrt{\mathcal{Q}^2+1} \left(2 c_1-c_5-2 c_6-8 c_7-4 c_8\right)
	\Big].
	\eea
	Note that the requirement $\Delta S_{\rm ext}\ge 0$ for all ${\cal Q}>0$ imposes inequality constraints \eqref{positivecond} on the coupling constants $c_i$.
	
	For large ${\cal Q}$, we have
	\bea
	M_{\mathrm{ext}}&=&\frac{\sigma  \ell _0}{3 \sqrt{6}}\left(1-
	\frac{6 \sigma }{5 \ell _0^2}(5 c_1+2 c_2+3 c_3+30 c_4+9 c_5+9 c_6+12 c_7+6 c_8)
	\right)\mathcal{Q}^{3/2}\cr
	&&
	+\frac{\sigma  \ell _0}{2 \sqrt{6}}\left(
	1-\frac{6 \sigma }{\ell _0^2}(3 c_1+c_2+c_3+2 c_4+c_5+c_6+4 c_7+2 c_8)
	\right)\mathcal{Q}^{1/2}\cr
	&&
	-\frac{\sigma  \ell _0}{8 \sqrt{6}}\left(
	1-\frac{6 \sigma }{\ell _0^2}(5 c_1-5 c_3-2 c_4-3 c_5-3 c_6-20 c_7-10 c_8)
	\right) \mathcal{Q}^{-1/2} + {\cal O}({\cal Q}^{-3/2}),\cr
	S_{\mathrm{ext}}&=&
	\frac{\pi   \sigma  \ell _0^2}{6} \left(1-\frac{6 \sigma }{\ell _0^2}(2 c_1+c_2+2 c_3+4 c_4+2 c_5+4 c_6+8 c_7+4 c_8)
	\right)\mathcal{Q}\cr
	&&-\frac{\pi  \sigma  \ell _0^2}{6}  \left(1-\frac{12 \sigma }{\ell _0^2}(2 c_1-c_5-2 c_6-8 c_7-4 c_8)
	\right)+\frac{\pi  \sigma  \ell _0^2}{12}\Big(1-\frac{6 \sigma }{\ell _0^2}(
	6 c_1+3 c_2\cr
	&&+6 c_3-4 c_4+2 c_5+4 c_6+24 c_7+12 c_8)
	\Big) \mathcal{Q}^{-1} + {\cal O}({\cal Q}^{-2}).
	\eea
	For small ${\cal Q}$, the leading-order expansions involve only the higher-derivative corrections, namely
	\bea
	M_{\mathrm{ext}}&=&-\frac{2 \sqrt{3} \sigma ^2}{5 \mathcal{Q} \ell _0}(c_2+4 c_3+2 c_5+2 c_6+16 c_7+8 c_8) +\frac{\mathcal{Q} \sigma  \ell _0}{2 \sqrt{3}}\Big[\cr
	&&1-\frac{3 \sigma }{2 \ell _0^2}(8 c_1+3 c_2+4 c_3+16 c_4+6 c_5+6 c_6+16 c_7+8 c_8)
	\Big] + {\cal O}({\cal Q}^2),\cr
	S_{\mathrm{ext}}&=&-2 \pi  \sigma ^2(c_2+2 c_3+2 c_5+4 c_6+16 c_7+8 c_8)\cr
	&&+\frac{ \pi  \mathcal{Q}^2 \sigma  \ell _0^2}{12}\Big[
	1-\frac{12 \sigma }{\ell _0^2}(4 c_4+c_5+2 c_6)
	\Big] + {\cal O}({\cal Q}^3)\,.
	\eea

\end{document}